\documentclass{emulateapj}

\usepackage{graphicx}
\usepackage[space]{grffile}
\usepackage{latexsym}
\usepackage{amsfonts,amsmath,amssymb}
\usepackage{url}
\usepackage[utf8]{inputenc}
\usepackage{hyperref}
\hypersetup{colorlinks=false,pdfborder={0 0 0}}
\usepackage{textcomp}
\usepackage{longtable}
\usepackage{multirow,booktabs}
\usepackage[english]{babel}
\usepackage{natbib}

\usepackage{natbib}

\def \TI {$^{44}$Ti}

\shorttitle{The distribution of radioactive $^{44}$Ti in Cassiopeia A}
\shortauthors{Grefenstette et al.}

\begin{document}


\title{The distribution of radioactive $^{44}$Ti in Cassiopeia A}


\author{
Brian W.~Grefenstette\altaffilmark{1},
Chris L.~Fryer\altaffilmark{2},
Fiona A.~Harrison\altaffilmark{1},
Steven E.~Boggs\altaffilmark{3}, % responded
Tracey DeLaney\altaffilmark{4},
J. Martin Laming\altaffilmark{5}, % Will check back after the weekend
Stephen P.~Reynolds\altaffilmark{6},
David M.~Alexander\altaffilmark{7}, % responded with affiliation change
Didier Barret\altaffilmark{8},
Finn E. Christensen\altaffilmark{9},
William W. Craig\altaffilmark{3,10},
Karl Forster\altaffilmark{1},
Paolo Giommi\altaffilmark{11},
Charles J. Hailey\altaffilmark{12},
Alan~Hornstrup\altaffilmark{9},
%Victoria M.~Kaspi\altaffilmark{13}, % Asked to be removed
Takao Kitaguchi\altaffilmark{13},
J. E. Koglin\altaffilmark{14},
Laura Lopez\altaffilmark{15},
Peter H. Mao\altaffilmark{1},
Kristin K.~Madsen\altaffilmark{1},
Hiromasa Miyasaka\altaffilmark{1},
Kaya Mori\altaffilmark{12},
Matteo Perri\altaffilmark{11,16},
Michael J.~Pivovaroff\altaffilmark{10}, % responded with commments
Simonetta Puccetti\altaffilmark{11,16}, 
Vikram Rana\altaffilmark{1},
Daniel Stern\altaffilmark{17},
Niels J.~Westergaard\altaffilmark{9},
Daniel R.~Wik\altaffilmark{18,19}, % Signed off
William W. Zhang\altaffilmark{19},
and Andreas~Zoglauer\altaffilmark{3}
}

\affil{
$^1${Cahill Center for Astrophysics, 1216 E. California Blvd, California Institute of Technology, Pasadena, CA 91125, USA}  \\
$^2${CCS-2, Los Alamos National Laboratory, Los Alamos, NM 87545, USA} \\
$^3${Space Sciences Laboratory, University of California, Berkeley, CA 94720, USA} \\
$^4${Physics \& Engineering Department, West Virginia Wesleyan College, Buckhannon, WV 26201, USA} \\
$^5${Space Science Division, Naval Research Laboratory, Code 7684, Washington, DC 20375, USA} \\
$^6${Physics Department, NC State University, Raleigh, NC 27695, USA} \\ 
$^7${Centre for Extragalactic Astronomy, Department of Physics, Durham University, Durham, DH1 3LE, U.K.} \\
$^8${Universite de Toulouse; UPS-OMP; IRAP; Toulouse, France \& CNRS; Institut de Recherche en Astrophysique et Plan\'etologie; 9 Av. colonel Roche, BP 44346, F-31028 Toulouse cedex 4, France} \\
$^{9}${DTU Space - National Space Institute, Technical University of Denmark, Elektrovej 327, 2800 Lyngby, Denmark} \\
$^{10}${Lawrence Livermore National Laboratory, Livermore, CA 94550, USA} \\
$^{11}${ASI Science Data Center (ASDC), via del Politecnico, I-00133 Rome, Italy} \\
$^{12}${Columbia Astrophysics Laboratory, Columbia University, New York, NY 10027, USA} \\
%$^{13}${Department of Physics, McGill University, Rutherford Physics Building, Montreal, Quebec H3A 2T8, Canada.}\\
$^{13}${Department of Physical Science, Hiroshima University, 1-3-1 Kagamiyama, Higashi-Hiroshima, Hiroshima 739-8526, Japan}\\
$^{14}${Kavli Institute for Particle Astrophysics and Cosmology, SLAC National Accelerator Laboratory, Menlo Park, CA 94025, USA} \\
$^{15}${Department of Astronomy and Center for Cosmology \& Astro-Particle Physics, The Ohio State University, Columbus, Ohio 43210}\\
$^{16}${INAF - Astronomico di Roma, via di Frascati 33, I-00040 Monteporzio, Italy}\\
$^{17}${Jet Propulsion Laboratory, California Institute of Technology, 4800 Oak Grove Drive, Pasadena, CA 91109, USA} \\
$^{18}${Department of Physics and Astronomy, Johns Hopkins University, Baltimore, MD 21218, USA}\\
$^{19}${NASA Goddard Space Flight Center, Greenbelt, MD 20771, USA} 
}
\email{bwgref@srl.caltech.edu}

\bibliographystyle{apj}

\begin{abstract}

The distribution of elements produced in the inner-most layers of a supernova explosion is a key diagnostic for studying the collapse of massive stars. Here we present the results of a 2.4 Ms \textit{NuSTAR} observing campaign aimed at studying the supernova remnant Cassiopeia A (Cas A). We perform spatially-resolved spectroscopic analyses of the $^{44}$Ti ejecta which we use to determine the Doppler shift and thus the three-dimensional (3D) velocities of the $^{44}$Ti ejecta. We find an initial $^{44}$Ti mass of 1.54 $\pm$ 0.21 $\times 10^{-4}$ M$_{\odot}$ which has a present day average momentum direction of 340$^{\circ}$ $\pm$ 15$^{\circ}$ projected on to the plane of the sky (measured clockwise from Celestial North)  and tilted by 58$^{\circ}$ $\pm$ 20$^{\circ}$ into the plane of the sky away from the observer, roughly opposite to the inferred direction of motion of the central compact object. We find some $^{44}$Ti ejecta that are clearly interior to the reverse shock and some that are clearly exterior to the reverse shock. Where we observe $^{44}$Ti ejecta exterior to the reverse shock we also see shock-heated iron; however, there are regions where we see iron but do not observe $^{44}$Ti. This suggests that the local conditions of the supernova shock during explosive nucleosynthesis varied enough to suppress the production of $^{44}$Ti in some regions by at least a factor of two, even in regions that are assumed to be the result of processes like $\alpha$-rich freezeout that should produce both iron and titanium. \\ 
  
\end{abstract}  


\section{Introduction}
\label{section:intro}

Young supernova remnants (SNR) are laboratories that we can use to study nucleosynthesis and the dynamics in supernova explosions. One key diagnostic in young remnants is the relative production of titanium, nickel, and silicon as observed through atomic transitions in the 0.1--10 keV band and, in the case of $^{44}$Ti, through $\gamma$-rays emitted through radioactive decay.

The physical processes that produce these elements in core-collapse supernova explosions depend on the local conditions of the shock during explosive nucleosynthesis. The innermost ejecta in the supernova, the silicon layer, is shock-heated, fusing the silicon into nickel. However, due to Coulomb repulsion, it is unlikely that two $^{28}$Si nuclei will fuse directly to $^{56}$Ni. Instead, photodisintegration first rearranges the abundances, producing a set of nuclei clusters with many reactions contributing to the final relic abundances.

The local supernova shock conditions can be parameterized by the peak temperature of the ejecta and the density of the ejecta at the peak temperature. Different regions of this parameter space correspond to different reactions dominating the nucleosynthesis \citep{Magkotsios:2010kr}. An example of this is a region where incomplete silicon photo-disintegration leads to partial silicon burning, leaving behind silicon-rich ejecta. Another is a region where a high density of free $\alpha$-particles results in the ``freezing out" of nuclear reactions \citep[i.e., ``$\alpha$-rich freezeout",][]{Woosley:1973ge} which can lead to ejecta rich with elements heavier than the iron group \citep[e.g.,][]{Woosley:1992dh}.

For a given supernova, the innermost ejecta have a wide range of peak temperatures and densities and, therefore, a wide range of reactions can play a role in the nucleosynthetic yields. The $^{44}$Ti yield is very sensitive to these conditions and thus is an ideal probe of nuclear and explosion physics. $^{56}$Ni is much less sensitive to these conditions and instead is ideally suited to delineating the region where silicon burning occurs.  Using the yields of both nickel and titanium we can better identify the burning regions and their exact conditions.

The abundance of $^{56}$Ni is measured by observing atomic transitions in shock-heated iron that are current present in the supernova remnant as iron is a decay product of $^{56}$Ni. Determining the exact iron abundance from observations is not, however, straightforward as this requires models of the shock heating and the ionization state of the iron and, crucially, requires a model of the density of the iron \citep[e.g.,][]{Hwang:2012gi}. In addition, some of the observed iron could be material swept up in the supernova shock rather than iron synthesized in the explosion. All these uncertainties make it difficult to determine the exact iron abundance and, therefore, the initial nickel abundance.

The abundance of $^{44}$Ti is easier to determine since it is seen via the radioactive decay of $^{44}$Ti$\rightarrow$$^{44}$Sc, producing a gamma-ray line at 1157 keV, and $^{44}$Sc$\rightarrow$$^{44}$Ca which produces a pair of gamma-ray lines at 78.32 and 67.87 keV. The branching ratios of the 1157, 78.32, and 67.87 keV lines are 99.9\%, 96.4\%, and 93\%, respectively \citep{Chen_2011}\footnote{We note that our branching ratios are based on the most recent nuclear physics measurements and are subtly different than those reported in other papers on $^{44}$Ti \citep[e.g.,][]{Grebenev:2012ks}.}. The present-day flux of photons produced in the radioactive decay of $^{44}$Ti is therefore directly proportional to the initial synthesized $^{44}$Ti mass, independent of local conditions. For SNR that are a few hundred years old, the $^{44}$Ti, which has a half-life of 58.9 $\pm$ 0.3 yr \citep{Ahmad:2006ba}, is still abundant enough to be observed.
\begin{figure}
\begin{center}
\includegraphics[width=1.0\columnwidth]{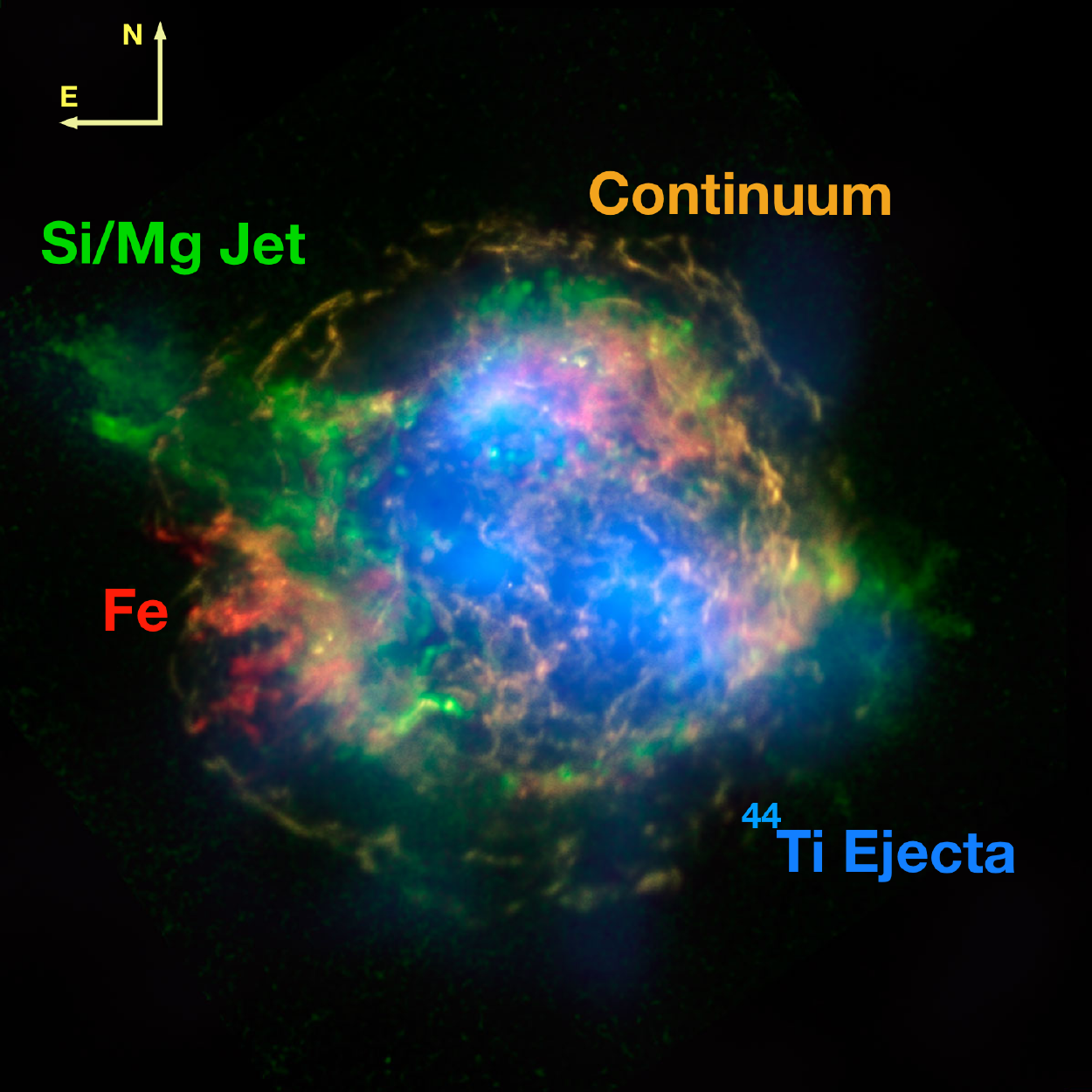}
\caption{\label{fig:beauty_shot}
The spatial distribution of $^{44}$Ti in Cas A compared with the other bright X-ray features. The \textit{NuSTAR} 65--70 keV background subtracted image covering the 68 keV $^{44}$Sc line tracing the $^{44}$Ti-rich ejecta is shown in blue. The \textit{NuSTAR} image has been adaptively smoothed for clarity. The 4--6 keV continuum observed by \textit{Chandra} image is shown in gold, the ratio in the Si/Mg band highlighting the NE/SW jet is shown in green (data courtesy NASA/CXC, Si/Mg ratio image J. Vink), while the distribution of X-ray emitting iron is shown in red (Fe distribution courtesy U. Hwang). Image credit, Robert Hurt, NASA/JPL-Caltech.
}
\end{center}
\end{figure}

Cassiopeia A (Cas A) is arguably the best-studied core-collapse SNR. It is young, with an explosion date inferred from the dynamical motion of the ejecta knots of 1671 \citep{Thorstensen:2001dza}, and relatively nearby at 3.4 kpc \citep{Reed:1995ks}.  $^{44}$Ti has been detected in Cas A by the COMPTEL instrument on the \textit{Compton Gamma-Ray Observatory (CGRO)} \citep{Iyudin_1994}, \textit{Beppo-SAX} \citep{Vink:2001bfa}, the IBIS/ISGRI instrument on the \textit{INTEGRAL} Observatory \citep{Renaud:2006grc}, and \textit{NuSTAR} \citep{Grefenstette:2014ds}. Upper limits consistent with the detections were obtained using the OSSE instrument on \textit{CGRO} \citep{The:1996},  \textit{RXTE} \citep{Rothschild:2003ev}, and the SPI instrument on \textit{INTEGRAL} \citep{Martin:2009}.  A comparison of the total yield from all of these observations demonstrates that all measurements of the initial $^{44}$Ti mass using the 68 and 78 keV lines are consistent with an initial $^{44}$Ti mass of 1.37 $\pm$ 0.19 $\times$ 10$^{-4}$ M$_{\odot}$ \citep{Siegert:2015fr}. However, only \textit{NuSTAR} is capable of spatially resolving the remnant and has the energy resolution to search for Doppler broadening of the 67.87 and 78.32 keV $^{44}$Ti decay lines.

Our analysis of the initial \textit{NuSTAR} observations demonstrated that the $^{44}$Ti is highly asymmetric and does not trace the observed distribution of Fe-K emission observed by \textit{Chandra} \citep[][and Figure \ref{fig:beauty_shot}]{Grefenstette:2014ds}.  A highly collimated axisymmetric jet engine had previously been invoked to explain the high ratio of $^{44}$Ti / $^{56}$Ni in Cas A \citep[e.g.,][]{Nagataki:1998jd}. However, the $^{44}$Ti ejecta does not appear to be collimated in a jet-like structure associated with the NE/SW Si/Mg jet (e.g. the green layer in Figure \ref{fig:beauty_shot}) observed by \textit{Chandra}, arguing that the Si/Mg asymmetric emission is not, in fact, indicative of a jet-driven explosion.

Unlike in SN1987A, where the $^{44}$Ti decay lines appear to be red-shifted but narrow \citep{Boggs_2015}, in Cas A we found that the decay lines were measurably broadened, indicating that there is a diversity in the direction of motion of the $^{44}$Ti ejecta. Our previous work using $\sim$ 1 Ms of {\em NuSTAR} observations did not have sufficient statistical power to perform a spatially resolved spectroscopic analysis of the $^{44}$Ti ejecta and so we were only able to describe the spatially integrated kinematics. 
\begin{figure*}
\begin{center}
\includegraphics[width=0.67\columnwidth]{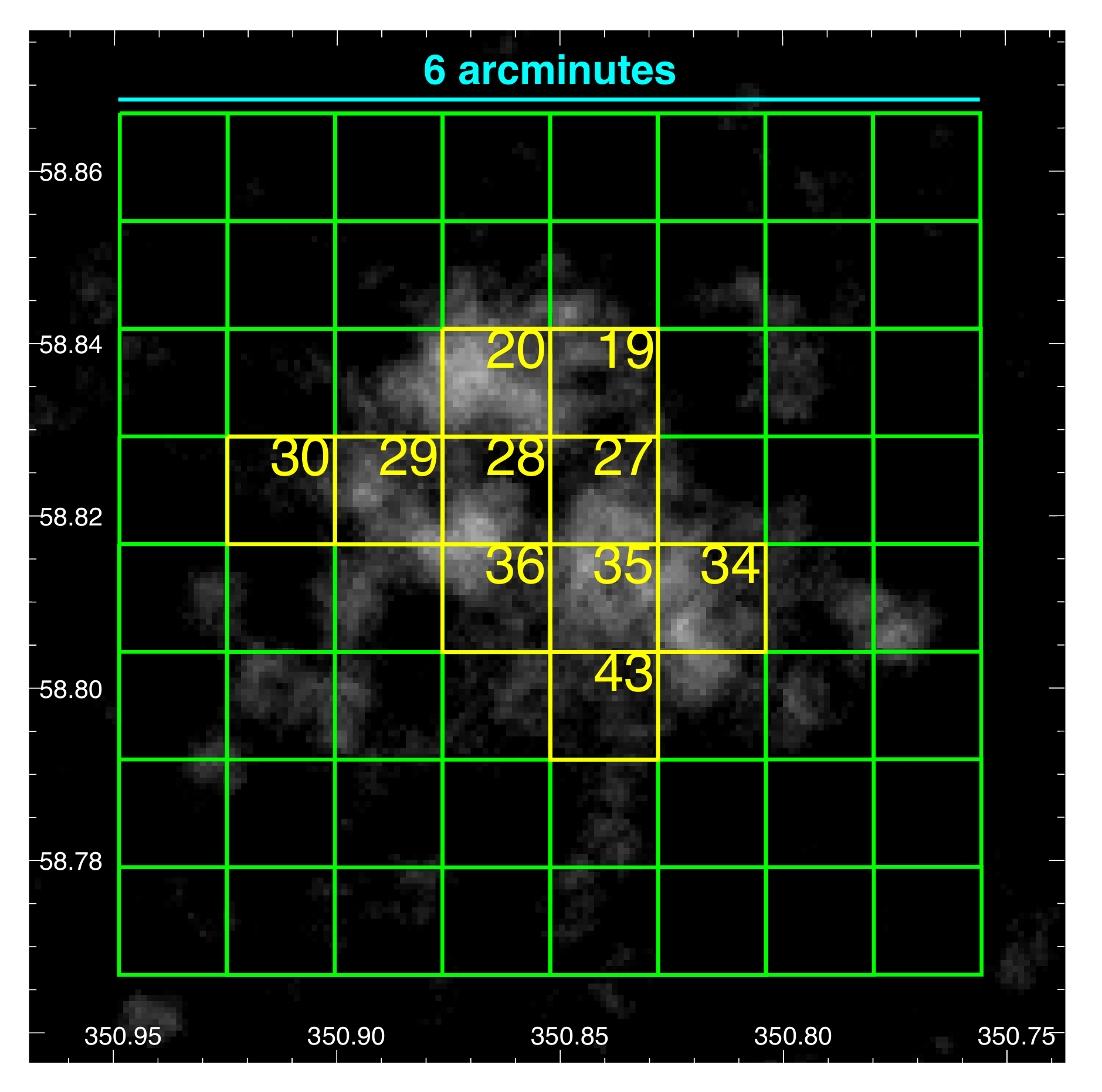} \includegraphics[width=0.67\columnwidth]{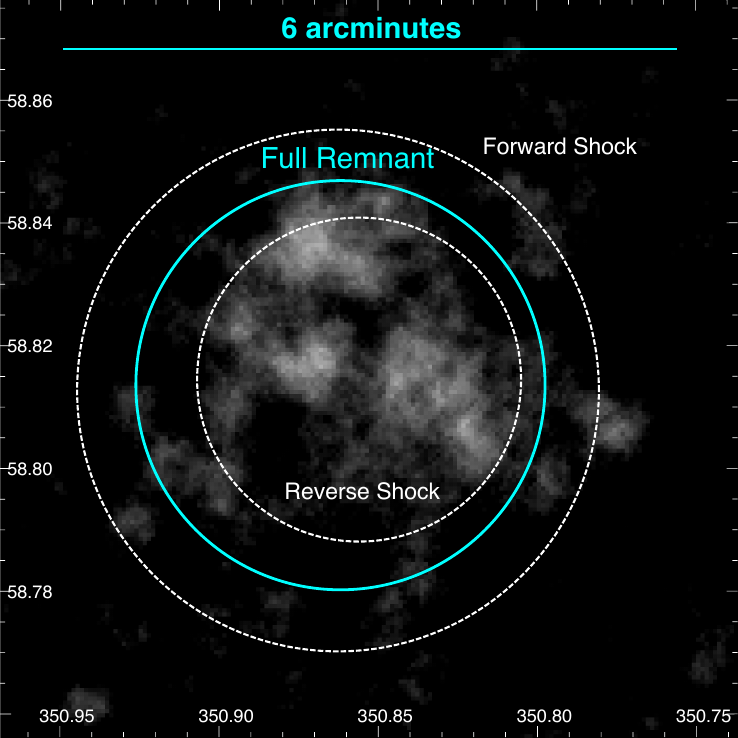}
\includegraphics[width=0.67\columnwidth]{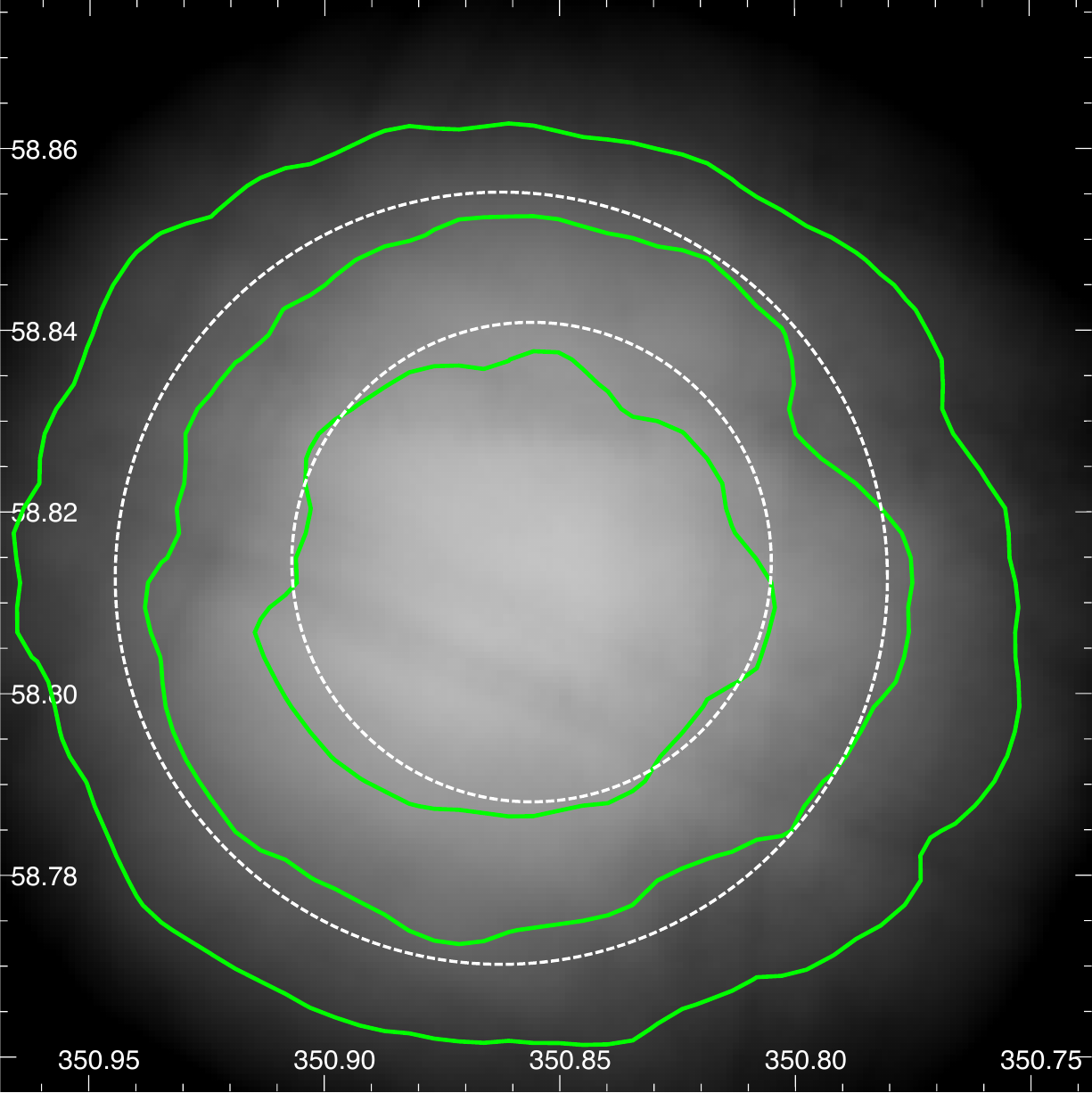}

\caption{\label{fig:TIboxes}
(\textit{Left}) The 65--70 keV background-subtracted {\em NuSTAR} image along with the regions used for spectral analysis. The data have been smoothed with a 7-pixel ($\sim$ 18$''$) ``top-hat" kernel for visual presentation. The green grid shows the regions used for spectral analysis with box 0 at top right and the box number increasing to the east (left). Box 8 is one row below box 0, Box 16 two rows below Box 0, and so on. The location of the grid was chosen to evenly cover the $^{44}$Ti. Yellow boxes with associated region numbers indicate where $^{44}$Ti has been detected in this analysis (see text for details). (\textit{Center}) The same data, but showing the region used for the integrated $^{44}$Ti flux estimate, which has a radius of 120$''$ and is centered on the remnant. This region is the same as used in \cite{Grefenstette:2014ds}. The location of the forward and reverse shocks as measured by \textit{Chandra} \citep{Gotthelf:2001bn} are shown by the white dashed circles. (\textit{Right}) The combined exposure map computed at 68 keV after integrating over all epochs and combining both telescopes is shown by the greyscale image with a linear stretch. The green contours show the location where the exposure has dropped to 90, 80, and 70\% of the maximum exposure, moving outward from the center.
}
\end{center}
\end{figure*}

In this paper, we present an analysis of 2.4 Ms of \textit{NuSTAR} observations. In \S \ref{section:obs} we describe the observations and the analysis techniques used to determine the 3-dimensional (3D) spatial positions of the $^{44}$Ti ejecta knots. In \S \ref{section:results} we present our results while in \S \ref{section:discussion} we compare and contrast the properties of the $^{44}$Ti with other known features of the remnant in the X-ray, infrared, and optical, and discuss the implications of these results in the context of theoretical models of the supernova explosion.

\section{Data and Methods}
\label{section:obs}

\subsection{{\em NuSTAR}  Data}
{\it NuSTAR} is the first focusing hard X-ray observatory. It is composed of two co-aligned X-ray telescopes (FPMA and FPMB) observing the sky in an energy range from 3--79 keV \citep{Harrison_2013}. The field of view of each {\em NuSTAR} telescope is roughly 12$'$ x 12$'$ and has a point-spread function (PSF) with a full-width, half maximum (FWHM) of 18$''$ and a half-power diameter of 58$''$. 

{\it NuSTAR} observed Cas A during the first 18 months of the {\it NuSTAR} mission (Table \ref{tab:obs}) with a total of 2.4 Ms of exposure time. These observations include the original $\sim$1 Ms of data that we have presented previously \citep{Grefenstette:2014ds}. We reduced the {\it NuSTAR} data with the {\it NuSTAR} Data Analysis Software (NuSTARDAS) version 1.4.1 and {\it NuSTAR} calibration database (CALDB) version 20150316 to produce images, exposure maps, and response files for each telescope.

We examined the background reports from the \textit{NuSTAR} SOC and identify solar flares during three sequence IDs (40021003003, 40021011002, and 40021015002). As the flux from the solar flares only affects the spectrum below 10 keV \citep{Wik:2014gg} and the amount of time that is affected by solar flares is small compared to the duration of the observations we do not apply any filtering to the data. We do not use the data for sequence IDs 40021001004, 40021002010, and 40021003002 as these short observations were significantly offset from the target pointing position.

\subsection{Data Reduction}

We leverage the increased exposure time relative our previous work to perform an analysis on smaller spatial scales, though at lower signal-to-noise than when integrating over the remnant as a whole.

Figure \ref{fig:TIboxes} shows the 65--70 keV \textit{NuSTAR} image of Cas A. To produce this image we combine all of the data (for all epochs and both telescopes) using \texttt{ximage} and then subtract the similarly combined background images. We smooth the result with an $\approx$18$''$ top-hat kernel to generate the underlying grey-scale images in the left and center panels. This is comparable to the image that we used for the analysis in \cite{Grefenstette:2014ds}. While the band image is useful, it can contain some contamination from the strong non-thermal emission that is present in the remnant \citep{Grefenstette:2015kz} and is also not optimized to search for emission that is red or blue-shifted where some of the line flux may fall outside of the 65--70 keV bandpass.

Instead of using the band image, we instead perform a systematic, spatially-resolved spectroscopic analysis by dividing the remnant into an 8$\times$8 grid of  regions (Figure \ref{fig:TIboxes}). Each grid box is a square with 45$''$ sides. The grid is centered by eye on the spatial distribution of the \TI~ ejecta. The right panel in Figure \ref{fig:TIboxes} shows the exposure maps computed at 68 keV (this accounts for the energy-dependent vignetting) and combined in the same way as the 65--70 keV band image. It demonstrates that the exposure is relatively uniform across the grid, only dropping by $\approx$30\% near the corners of the grid.

We use the \texttt{nuproducts} FTOOL to extract source spectra and generate ancillary response files (ARFs), which describe the effective area of the optics, as well as response matrix files (RMFs), which describe the response of the detectors. We generate simulated background spectra using \texttt{nuskybgd} \citep{Wik:2014gg} following the procedure described in \cite{Grefenstette:2014ds}. This results in 22 sets of data files (11 epochs $\times$ 2 telescopes) for each region. The ARF and RMF files computed by \texttt{nuproducts} account for the variations in the effective exposure due to vignetting described above.

We integrate over all 11 epochs by using the \texttt{addspec} FTOOL, setting \texttt{bexpscale=1} when calling \texttt{addspec} to prevent overflowing the exposure keyword. This results a two sets of source, background, ARF, and RMF files (one for FPMA and one for FPMB) for each region. \\

Since the $^{44}$Ti emission is extended (with a spatial distribution that we do not know a priori), we have to make a decision on how to normalize the ARF.

For the spectral analysis of point sources, \texttt{nuproducts} adjusts the normalization of the ARF (and thus the measured flux) to account for the fraction of the PSF that falls outside of the source region. This ``PSF correction" is not performed when observing extended sources because the correction assumes that the extraction region is precisely centered on the point source. Here the source flux is smeared out over the source extraction region, making an accurate PSF correction impossible. Instead we opt to simply apply no PSF correction to the ARF as the most conservative approach. We address the impact of this on the interpretation of the measured flux below.

\begin{figure}[ht]
\begin{center}
\includegraphics[width=\columnwidth]{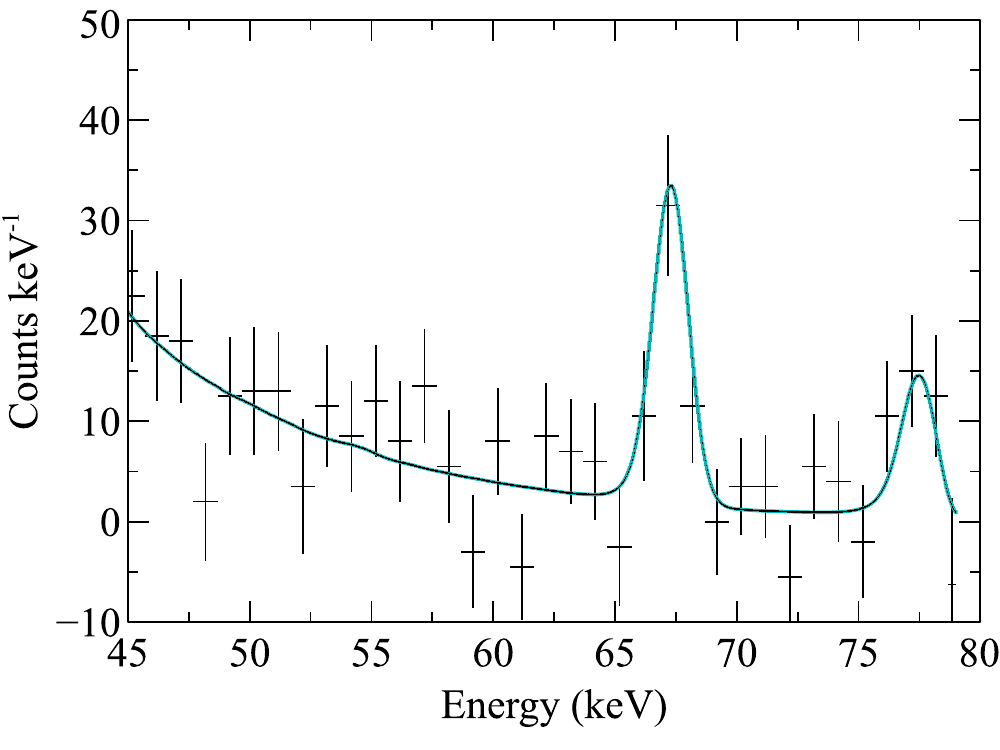}\\
\includegraphics[width=\columnwidth]{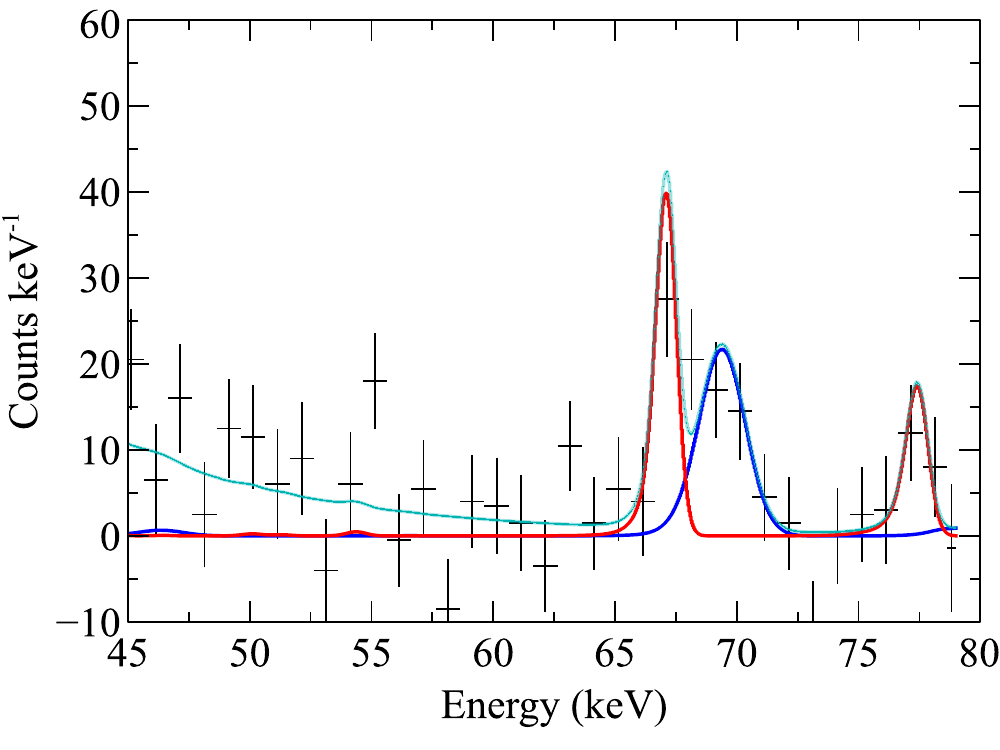}
\caption{
Spectral fits to two of the spectral regions. The data (black) are shown with 1-$\sigma$ error bars along with the best-fit line+continuum model (cyan). The data for the two telescopes are combined and re-binned for plotting purposes, while the models are shown un-binned. {\em Top}: Region 35, which has a marginally broadened pair of $^{44}$Ti lines. {\em Bottom}: Region 20, showing the narrow red-shifted component (for which both the 68 and 78 keV lines are observed) and the broad blue-shifted component (for which only the 68 keV line is in the {\em NuSTAR} bandpass). 
\label{fig:spec}
}
\end{center}

\end{figure}

\begin{table}
\begin{center}

      \caption{{\it NuSTAR} Observations of Cas A}
  \label{tab:obs}
    \begin{tabular}{lrc} \hline \hline              
  OBSID     	&	 Exposure   & 	UT Start Date    \\ \hline 
40001019002 &  294 ks & 2012 Aug 18 \\
40021001002 & 190 ks & 2012 Aug 27 \\
40021001004* & 29 ks &  2012 Oct 07 \\
40021001005 & 228 ks & 2012 Oct 07 \\
40021002002 & 288 ks & 2012 Nov 23 \\
40021002006 & 160 ks & 2013 Mar 02 \\
40021002008 & 226 ks & 2013 Mar 05 \\
40021002010* & 16 ks & 2013 Mar 09 \\
40021003002* & 13 ks & 2013 May 28 \\
40021003003 & 216 ks & 2013 May 28 \\
40021011002 & 246 ks & 2013 Oct 30 \\
40021012002 & 239 ks & 2013 Nov 27 \\
40021015002 & 86 ks & 2013 Dec 21 \\
40021015003 & 160 ks & 2013 Dec 23 \\
  \hline
  Total & $\approx$ 2.4 Ms & \\ \hline
    \end{tabular}
    \end{center}
*: Observations not considered here due to offsets from the desired pointing location.
\end{table}

\subsection{Spectral Fitting}

We performed spectral fitting using \texttt{XSPEC} \citep{Arnaud_1996} using the \texttt{cstat} statistic for the model fitting. In general, the observed (source+background) spectra satisfy the requirement that each bin contains at least one count, so we do not arbitrarily rebin the spectra before fitting. We simultaneously fit the spectra for each telescope, allowing a standard cross-normalization constant to account for variations in the overall effective area between the two telescopes.

The broadband hard X-ray spectrum of Cas A is dominated by thermal emission in the interior of the remnant along with a non-thermal tail throughout the remnant \citep[e.g.][]{Grefenstette:2015kz}. The non-thermal tail spatially varies across the remnant in both flux and spectral shape, so we fit each box with the \texttt{srcut} spectral model. We keep the radio spectral index fixed to 0.77 \citep[which is the radio spectral index integrated over the remnant,][]{Baars:1977va} and then fit for the break frequency and the normalization of the continuum. In the interior of Cas A, the data also require a thermal component, which we model using a simple \texttt{bremss} component. This thermal continuum can contribute significantly up to $\sim$15 keV. We mask the spectrum over the Fe-K line features in the 5--7 keV band in the {\em NuSTAR} spectrum. This results in a fit range of 3--4.5 keV and 8.5--79 keV.

To model the $^{44}$Ti decay lines we include two \texttt{gauss} components with the line widths and fluxes tied together and require that both lines have the some observed Doppler shift. The {\it NuSTAR} optics have an absorption edge at $\sim$78.4 keV \citep{Madsen:2015}, so the 78.32 keV decay line is only visible when the material is stationary or redshifted. Where the 67.87 keV line is blue-shifted we only fit with a single \texttt{gauss} component rather than the two components tied together. To determine whether the line is detected we compare the cash fit statistic with and without the $^{44}$Ti lines and require that the change in fit statistic is $>$9.0 to declare the $^{44}$Ti emission to be consistent with the data. We then use the \texttt{error} command to generate 90\% and 1-$\sigma$ confidence regions for the line centroid, Gaussian width, and line flux. Unless otherwise stated, all uncertainties quoted in the text are 90\% error estimates.

Overall, 10 of the 64 regions satisfied our detection conditions (the fit parameters are given in Table \ref{tab:fits}).

In most of these cases the measured Gaussian 1-$\sigma$ width of the lines are consistent with the energy resolution of the detectors \citep[$\approx$0.6 keV FWHM at 68 keV,][]{Harrison_2013}). This implies that the ejecta within each 45$''$ region samples $^{44}$Ti ejecta traveling in roughly the same direction or have a spread in velocities below the ability of \textit{NuSTAR} to detect (i.e. the top panel in Figure \ref{fig:spec}).

The one exception is region 20 (Figure \ref{fig:spec}, bottom panel), which contains line emission clearly broadened beyond the instrument response. In this case we added a second line and achieved an improved fit to the data, resulting in one red-redshifted component (20a) and a broad ($\sim$1 keV Gaussian width) blue-shifted component (20b). 

\subsection{Systematic Errors}

 Systematic errors in the detector gain calibration could influence the measured Doppler shift of the lines. The systematic uncertainties for {\em NuSTAR} are roughly $2\times10^{-4}$ in gain and 40 eV in offset \citep{Madsen:2015}. At 67.86 keV, the gain uncertainty yields a systematic error of 13 eV, or 60 km sec$^{-1}$, while the 40 eV offset uncertainty results in an systematic uncertainty in velocity of 180 km sec$^{-1}$. Both of these effects are significantly smaller than the statistical errors, so we neglect them below. \\

``Look-back" effects can change the apparent bulk location of the ejecta. This is entirely an effect of the light travel time difference between blue-shifted and red-shifted ejecta, where the red-shifted ejecta is ``younger" than the blue-shifted ejecta. For unresolved sources, this can cause spherically symmetric sources to appear red-shifted, especially for the gamma-ray lines from rapidly expanding supernovae \citep[e.g.][]{Chan:1988}. For Cas A, each region represents the integration over the line-of-sight distribution of the ejecta in the remnant. We can estimate the difference in flux due to look-back effects for ejecta. The maximum observed red-shifted line had a centroid of 67.13 keV (or 1\% $c$). This produces a 1 lyr line-of-sight offset per 100 yr. For a symmetric distribution of ejecta 340 yr after the explosion, the difference in apparent age between the front and rear extreme ejecta is 6.84 years. For an e-folding time of 86.54 yr, this results in a difference in observed flux due to look-back effect of only $\sim$8\%. We conclude that look-back effects do not significantly affect our results. 

\section{Results}
\label{section:results}

\begin{figure}
\begin{center}
\includegraphics[width=1\columnwidth]{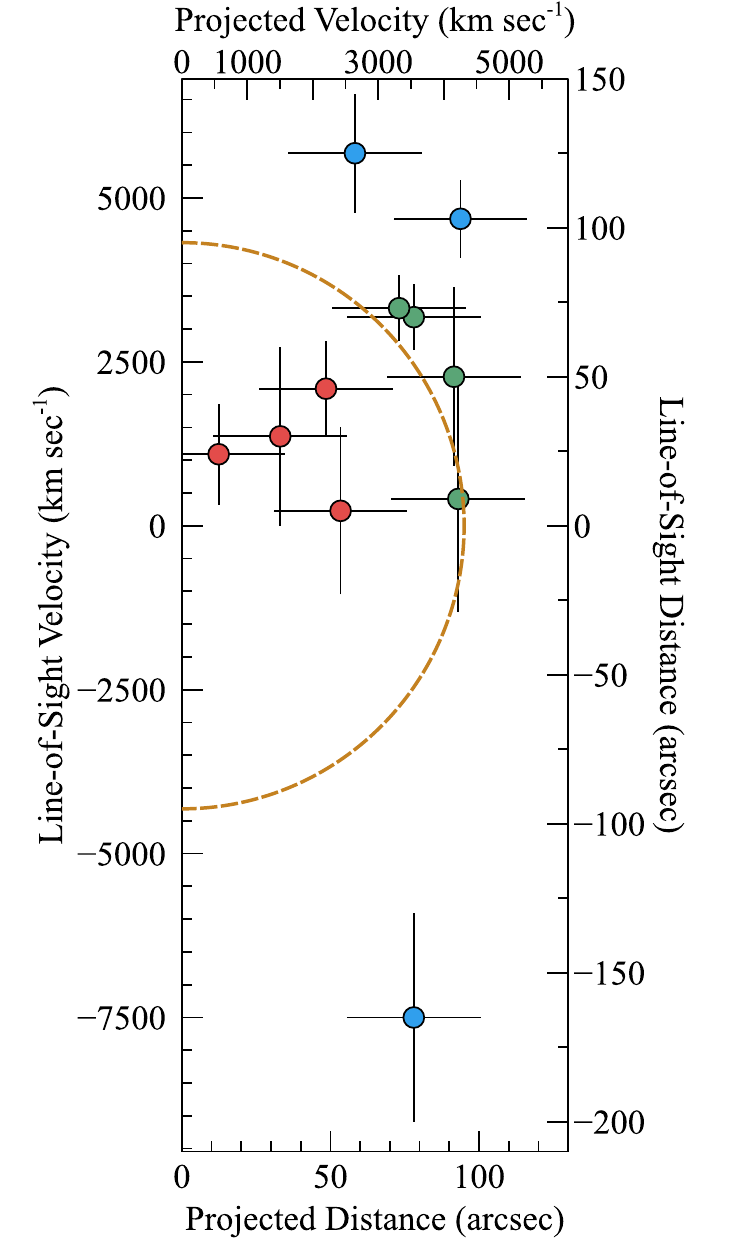}
\caption{\label{fig:prop_vs_los}
The measured projected distance (lower X-axis) and the measured velocity along the line-of-sight (left Y-axis) along with 1-$\sigma$ error bars. The regions that are more than 1-$\sigma$ interior to the reverse shock (here assumed to be a sphere with a radius of 95$''$ and represented by the dashed gold curve) are color coded in red, the regions that are near the shock radius are color coded green, while the regions more than 1-$\sigma$ exterior to the reverse shock are color coded in blue. The secondary axes gives the conversion to a velocity and distance assuming a proportionality constant of 0.022$''$ per km sec$^{-1}$ (see text for details). }
\end{center}
\end{figure}

\begin{figure*}
\begin{center}
\includegraphics[width=2\columnwidth]{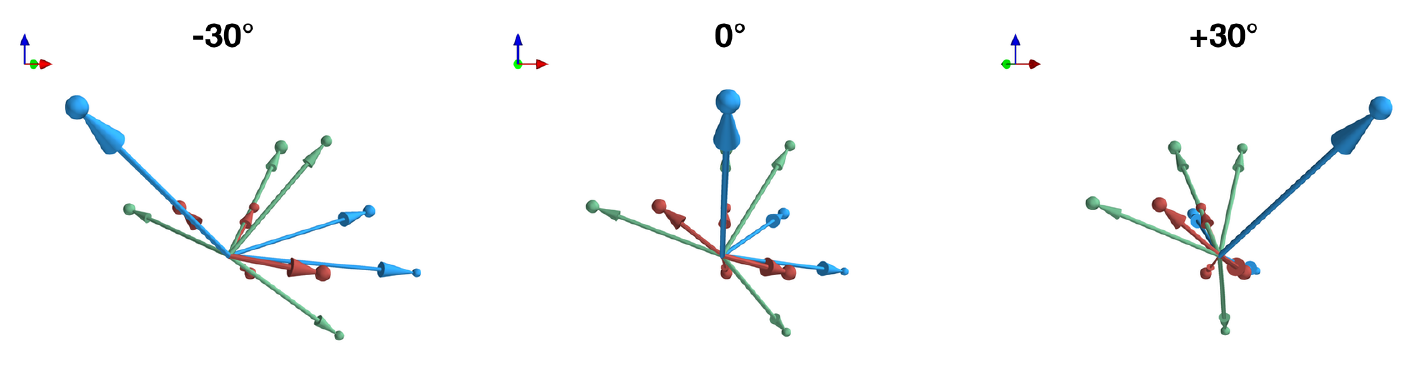}
\caption{\label{fig:vectors_3d_grid}
The 3D distribution of the $^{44}$Ti ejecta. The unit vectors are North (blue), West (red), and along the observer's line-of-sight (green). The data vectors have an origin at the center of expansion of the remnant. The color coding is the same as for Figure \ref{fig:prop_vs_los}. The center frame shows the remnant as seen by the observer, while the right/left frames have been rotated +/- 30$^{\circ}$ clockwise around North (blue) axis. An animation showing the full rotation of the remnant is available in the online journal.}
\end{center}
\end{figure*}

\begin{figure}
\begin{center}
\includegraphics[width=1\columnwidth]{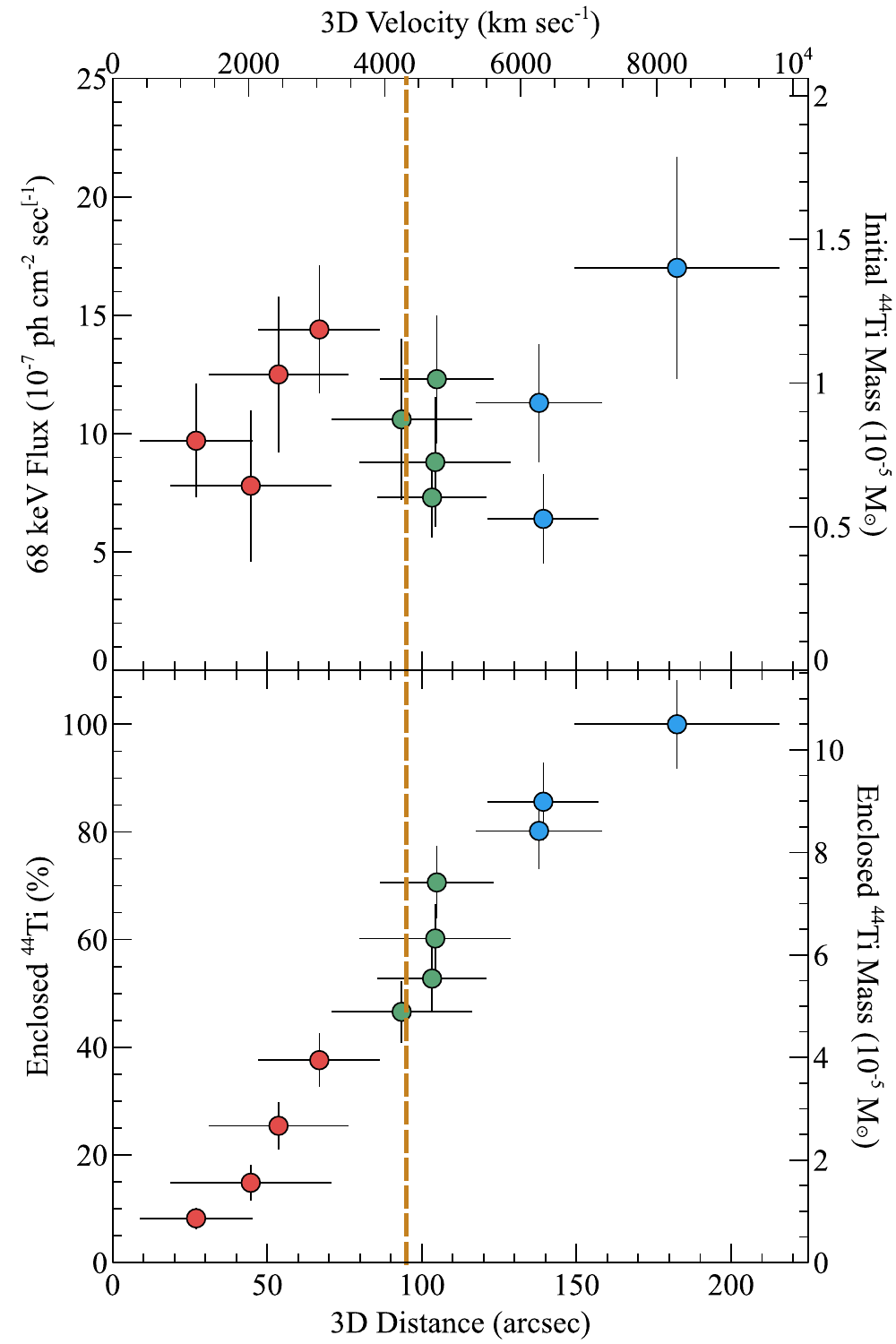}
\caption{\label{fig:space_velocity_vs_flux}
\textit{(Top)} The flux in the 68 keV line (Y-axis) for each region is plotted against the 3D space velocity of each region (X-axis). The secondary Y-axis (right) shows the inferred initial mass of $^{44}$Ti for each region assuming the distance and age of the remnant given in the text. \textit{(Bottom)} The enclosed flux/mass fraction as a function of radius. Roughly 40\% of the $^{44}$Ti is clearly interior to the reverse shock, while 40\% of the mass is at or near the reverse shock radius, leaving roughly 20\% of the mass clearly exterior to the reverse shock. The color coding is the same as for Figure \ref{fig:prop_vs_los}.
 }
\end{center}
\end{figure}

\subsection{The 3-Dimensional Distribution of $^{44}$Ti in Cas A}

We can combine the distance from the center of expansion of the remnant and the observed Doppler shift of the ejecta to determine the 3D velocity of each ejecta knot. We measure the projected offsets in the plane of the sky from the center of expansion of the remnant \citep[$\alpha$(J2000)=23h23m27.77s$\pm$0.05s, $\delta$(J2000)=58$^{\circ}$48$'$49.4$''\pm$0.4$''$,][]{Thorstensen:2001dza}. If we assume the material is freely expanding, then the observed velocity is proportional to the distance from the center of expansion of the remnant. \cite{DeLaney:2010fy} found a proportionality constant for undecelerated ejecta of 0.022$''$ per km sec$^{-1}$. However, if a knot of $^{44}$Ti ejecta has encountered the reverse shock, then the knot will be decelerated by some amount that will depend on the density of the knot and the local speed of the reverse shock at the time of the encounter. For simplicity, we will adopt the undecelerated proportionality constant for undecelerated ejecta for all of the $^{44}$Ti knots as this is the most conservative projection into 3D. 

Using this proportionality we can convert between observed offsets (in arcsec) and velocity (in km sec$^{-1}$). Figure \ref{fig:prop_vs_los} shows the projected distance in the plane of the sky for each region and the measured velocity along the line of sight along with a fiducial reverse shock radius of 95$''$, while Figure \ref{fig:vectors_3d_grid} and its associated animation show a proper 3D representation of the data. 

Nearly all the $^{44}$Ti ejecta are seen traveling away from the observer. However, unlike in SN1987A, where the $^{44}$Ti ejecta appear to be traveling in the same direction with the same velocity \citep{Boggs_2015}, here we see that the ejecta are expelled into a large solid angle. We also see ejecta that are traveling at different speeds in roughly the same direction. There is also evidence for high-velocity $^{44}$Ti ejecta that have passed beyond the reverse shock (Figure \ref{fig:space_velocity_vs_flux}). However, one of these knots is region 20b, which we cannot indisputably identify as a single coherent feature as it is broadened along the line of sight. Even so, there is clearly significantly blue-shifted emission in this region, so there must be some $^{44}$Ti ejecta that have been ejected beyond the reverse shock radius. The fact that these ejecta are beyond the reverse shock implies that our assumption that these ejecta are freely expanding is probably incorrect since they should have been decelerated as they traversed the reverse shock. However, the amount of deceleration the ejecta experience depends on the density of the ejecta and we have no way of measuring the density of the $^{44}$Ti ejecta knots. We therefore consider the positions along the line-of-sight to be lower limits for these data.

\begin{table*}
\begin{center}

      \caption{{\it NuSTAR}  Fits Results
      	  \label{tab:fits}
      	}
    \begin{tabular}{|lcc|cc|cc|ccl|} \hline
     & & & \multicolumn{2}{|c}{\texttt{bremss}} & \multicolumn{2}{|c|}{\texttt{srcut}, $\alpha$=0.77} & \multicolumn{3}{c|}{\texttt{gauss}} \\
              
  ID     	&	 R.A.  & Dec. & kT (keV) & Norm ($10^{-3}$) & Break ($10^{17}$ Hz) & Norm$^{a}$ &  Centroid (keV) & 1-$\sigma$ Width (keV) & Flux$^{b}$     \\
   \hline 
19 &   350.8401 &    58.8355 & 2.34 $\pm$ 0.06 & 7.3 $\pm$ 0.3 & 2.07 $\pm$ 0.09 & 90.9 $\pm$ 6.5 & 67.35 $\pm$  0.29 & $<$0.3 & 8.8 $\pm$  2.8 \\
20a &   350.8643 &    58.8355 &  2.14$_{-0.06}^{+0.15}$ & 8.3 $\pm$ 0.3 & 2.1$_{-0.1}^{+5.7}$ & 65$_{-20}^{+5}$ & 67.15 $\pm$ 0.2 & $<$0.3 & 12.3 $\pm$ 4.9 \\
20b$^{*}$ &   - &    - &  - & - & - & - & 69.5 $\pm$ 0.6 & 0.9$_{-0.4}^{+0.6}$ & 17.2$_{-7.4}^{+8.6}$ \\
27 &   350.8401 &    58.8230 &  2.26 $\pm$ 0.09 & 8.9 $\pm$ 0.3 & 2.8 $\pm$ 0.3 & 60 $\pm$ 10 & 66.6 $\pm$ 0.3 & 0.61$^{+0.4}_{-0.25}$  & 11.3$_{-3.8}^{+4.6}$ \\
28 &   350.8643 &    58.8230 & 2.1 $\pm$ 0.1 & 8.2 $\pm$ 0.4 & 2.7 $\pm$ 0.2 & 56.5 $\pm$ 8 & 67.6 $\pm$ 0.5 & $<$0.92 & 7.8 $\pm$ 5.0 \\
29 &   350.8884 &    58.8230 & 2.1 $\pm$ 0.2 & 6 $\pm$ 0.3 & 2.37$_{-0.48}^{+0.25}$ & 37$_{-5}^{+14}$ &  67.8$_{-0.9}^{+0.4}$ & 0.7 $\pm$ 0.3 & 12.5 $\pm$ 5.5 \\
30$^{*}$ &   350.9125 &    58.8230 & 2.3 $\pm$ 0.1 & 6.8 $\pm$ 0.3  & 3.3$_{-0.8}^{+0.5}$ & 16$_{-3}^{+9}$ & 67.8 $\pm$  0.6 & 0.73$_{-0.6}^{+0.4}$ & 10.6 $\pm$ 5.5 \\
34 &   350.8160 &    58.8105 & 2.5 $\pm$ 0.1 & 7.1 $\pm$ 0.3 & 3.5 $\pm$ 0.2 & 82 $\pm$ 5 & 66.8 $\pm$ 0.2 & $<$0.6 & 6.4$_{-2.5}^{+4.1}$ \\
35 &   350.8401 &    58.8105  & 2.4 $\pm$ 0.1 & 7 $\pm$ 0.3 & 3.5 $\pm$ 0.4 & 46$_{-4}^{+9}$ & 67.4 $\pm$ 0.3 & 0.64 $\pm$ 0.3 & 14.4 $\pm$ 4.5 \\
36 &   350.8643 &    58.8105 & 2.2 $\pm$ 0.1 & 7.4 $\pm$ 0.3 & 3.6 $\pm$ 0.2 & 48 $\pm$ 4 & 67.6 $\pm$ 0.3 & 0.43 $\pm$ 0.3 & 9.6 $\pm$ 4 \\
43 &   350.8401 &    58.7980 & 2.13 $\pm$ 0.1 & 5.9 $\pm$ 0.1 & 2.72$^{**}$ & 51$^{**}$ & 67.1 $\pm$ 0.2 & $<$0.34 & 7.3$_{-3.2}^{+1.9}$ \\
  \hline
 \end{tabular}
  \end{center}
%Uncertainties are 90\% error ranges. 
$a$: Flux at 1 GHz in Jy; $b: 10^{-7}$ ph cm$^{-2}$ sec$^{-1} $ ; 
$^{*}$: Only fit with a single Gaussian line ; 
$^{**}$: Not well constrained.  \\
\end{table*}

\begin{table*}
\begin{center}

      \caption{{\it NuSTAR} 3D Data}
  \label{tab:converted}
    \begin{tabular}{|c|ccc|ccc|} \hline 
    & \multicolumn{3}{|c|}{Offsets (arcsec)} & \multicolumn{3}{|c|}{Velocities (km sec$^{-1}$)} \\
 
   ID     	&	West$^{a}$  & North$^{a}$ & Line-of-Sight$^{b}$  & West  & North & Line-of-Sight $^{b}$ \\
   \hline 
19 & 48 $\pm$ 22.5 & 78 $\pm$ 22.5 & 50 $\pm$ 30 & 2200 $\pm$ 1020 & 3500 $\pm$ 1020 & 2300 $\pm$ 1400  \\
20a & 3 $\pm$ 22.5 & 78 $\pm$ 22.5 & 70 $\pm$ 11 & 140 $\pm$ 1020 & 3500 $\pm$ 1020 & 3200 $\pm$ 500  \\
20b & 3 $\pm$ 22.5 & 78 $\pm$ 22.5 & -170 $\pm$ 35 & 140 $\pm$ 1020 & 3500 $\pm$ 1020 & -7500 $\pm$ 1600  \\
27 & 48 $\pm$ 22.5 & 33 $\pm$ 22.5 & 125 $\pm$ 20 & 2200 $\pm$ 1020 & 1500 $\pm$ 1020 & 5700 $\pm$ 910  \\
28 & 3 $\pm$ 22.5 & 33 $\pm$ 22.5 & 30 $\pm$ 30 & 120 $\pm$ 1020 & 1500 $\pm$ 1020 & 1400 $\pm$ 1400 \\
29 & -42 $\pm$ 22.5 & 33 $\pm$ 22.5 & 5 $\pm$ 28 & -1900 $\pm$ 1020 & 1500 $\pm$ 1020 & 230 $\pm$ 1300  \\
30 & -87 $\pm$ 22.5 & 33 $\pm$ 22.5 & 9 $\pm$ 38 & -4000 $\pm$ 1020 & 1500 $\pm$ 1020 & 410 $\pm$ 1700 \\
34 & 93 $\pm$ 22.5 & -12 $\pm$ 22.5 & 100 $\pm$ 13 & 4200 $\pm$ 1020 & -550 $\pm$ 1020 & 4700 $\pm$ 590  \\
35 & 47 $\pm$ 22.5 & -12 $\pm$ 22.5 & 46 $\pm$ 16 & 2100 $\pm$ 1020 & -550 $\pm$ 1020 & 2100 $\pm$ 730  \\
36 & 3 $\pm$ 22.5 & -12 $\pm$ 22.5 & 24 $\pm$ 17 & 140 $\pm$ 1020 & -550 $\pm$ 1020 & 1100 $\pm$ 770 \\
43 & 47 $\pm$ 22.5 & -56 $\pm$ 22.5 & 73 $\pm$ 11 & 2100 $\pm$ 1020 & -2500 $\pm$ 1020 & 3300 $\pm$ 500 \\
\hline
Mean$^{c}$ & 10.8 $\pm$  8.0 & 30.4 $\pm$ 8.0 & 20.2 $\pm$ 11.0 & 490 $\pm$ 360 & 1380 $\pm$ 380 & 920 $\pm$ 510 \\
 \hline
 \end{tabular}
 \end{center}
a: West and North uncertainties are the half-length of the square regions. \\
b: Line-of-sight uncertainties are based on the 1-$\sigma$ error ranges from the line fits. \\
c: Flux-weighted mean of all regions. \\
Error bars are 1-$\sigma$ and include the uncertainties on the measured flux from Table \ref{tab:fits}.
\end{table*}
\begin{figure*}
\begin{center}
\includegraphics[width=2\columnwidth]{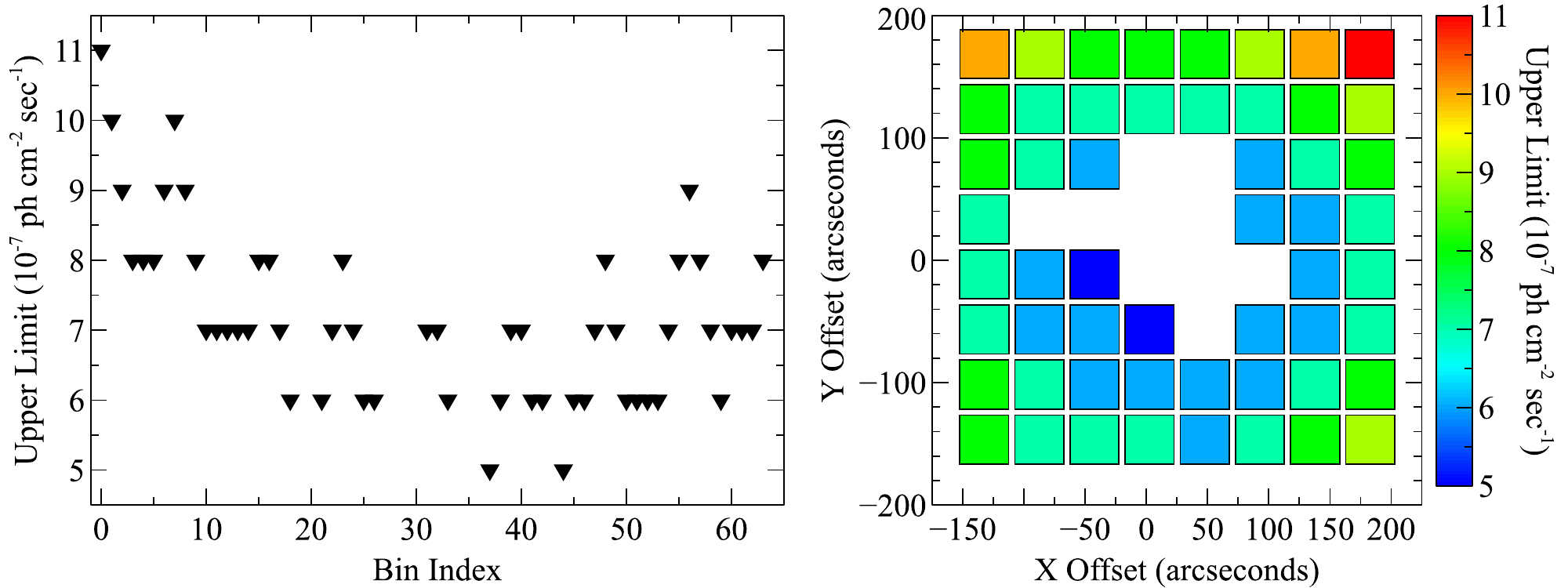}
\caption{\label{fig:upperlimits_50}
The upper limits for each region where the $^{44}$Ti is not detected ({\em left}) and the spatial distribution of the upper limits ({\em right}). Regions where the $^{44}$Ti is detected are intentionally left blank. See the text for details. }
\end{center}
\end{figure*}

\subsection{The Total Mass of $^{44}$Ti in Cas A Measured by {\em NuSTAR} }

Simply combining the fluxes from the detected regions in Table \ref{tab:fits} does not result in a good estimate of the total flux measured by {\em NuSTAR}.

As described above, for point sources a PSF correction is applied to account for the fraction of the \textit{NuSTAR} PSF that falls outside of the extraction radius. For extended sources it's not possible to apply an accurate PSF correction (and thus correctly normalize the flux) when combining neighboring regions as a region may contribute flux to its neighbors and correcting for this ``loss" will over-correct the flux.

This can be avoided by simply integrating over a larger region that covers all of the $^{44}$Ti emission. Here we compute the total flux by integrating over a 120$''$-radius extraction region centered on the remnant (Figure \ref{fig:TIboxes}, center panel). The standard behavior of \texttt{nuproducts} when producing extended ARFs is to assume that the spatial distribution of the counts for an extended source is described by the observed distribution of counts over a given energy range (i.e., that the spatial distribution of low-energy counts is the same as the distribution of high-energy counts). Here, since the source is background-dominated and since the $^{44}$Ti emission does not follow any other energy band where the images are source-dominated, we instead use the \texttt{flatflagarf} keyword to produce the ARFs. This explicitly assumes the prior of a ``spatially flat" distribution of source flux across the 120$''$ source region. We note that this option was not available for the previous analysis.

We fit the data with a power-law continuum and a single Gaussian line (fitting over the 10--72 keV band) and a power-law continuum with two Gaussian lines with the Doppler-shift of the lines, the line width, and line flux tied together (fitting over the 10--80 keV band), though the choice of model does significantly affect our results (Table \ref{tab:intfits}).

We find a 68 keV line flux that is slightly higher than we previously reported (1.84 $\pm$ 0.25 $\times 10^{-5}$ ph~sec$^{-1}$~cm$^{-2}$ compared with 1.53 $\pm$ 0.31 $\times 10^{-5}$ ph~sec$^{-1}$~cm$^{-2}$), though the line centroid(s) and line width(s) are both consistent with our previous results. The change in flux likely arises from the improvements in the generation of the ARF (we consider the new method to be superior). Taking a distance of 3.4 kpc, an explosion date of 1671, and an average epoch of the observations of 2013 gives a total initial mass of $^{44}$Ti of 1.54 $\pm$ 0.21 $\times 10^{-4}$ M$_{\odot}$ (compared with our previous value of 1.25 $\pm$ 0.30 $\times 10^{-4}$ M$_{\odot}$). 

\begin{table*}
	\begin{center}
		\caption{{\it NuSTAR}  Cas A Integrated Fits   \label{tab:intfits} }
		\begin{tabular}{|c|cc|ccc|c|} \hline
   & \multicolumn{2}{|c|}{\texttt{power-law} Parameters} & \multicolumn{3}{|c|}{67.87 keV \texttt{gauss} Parameters} & Initial $^{44}$Ti  \\
\hline
              
			 Line Model (Fit Range)  &	 $\Gamma$ & Norm$^{a}$ & Centroid$^{b}$ (keV) & Width$^{b}$ & Flux$^{c}$ &   Mass$^{d}$ \\
			Single line (10--72 keV) & 3.36 $\pm$ 0.05 & 1.229 $\pm$ 0.015 & 67.44$^{+0.11}_{-0.14}$ & 0.68 $\pm$ 0.15  & 1.84 $\pm$ 0.25 & 1.52 $\pm$ 0.2 \\
			Two Lines (10--80 keV) & 3.36 $\pm$ 0.05 & 1.229 $\pm$ 0.015 & 67.41$^{+0.10}_{-0.12}$ & 0.69 $\pm$ 0.15 & 1.87 $\pm$ 0.24 & 1.54 $\pm$ 0.2 \\
			\hline
			%Total & & & & 28.5 $\pm$ 3.5 \\
		\end{tabular}
	\end{center}
%	Uncertainties are 90\% error ranges. \\
	$
	a: \textrm{Flux at }1~\textrm{keV}, $ph cm$^{-2}$ sec$^{-1} $ ; 
	b: keV; 
	c: $10^{-5}$ ph cm$^{-2}$ sec$^{-1} $;
	d: $10^{-4}$ M$_{\odot}$
\end{table*}

\subsection{Flux Upper Limits}

In regions where we do not detect $^{44}$Ti emission we instead define the upper limit to be ``the flux at which 50\% of the time we would have detected the $^{44}$Ti". We determine this by repeatedly simulating the source and background spectra using a power-law continuum (fit to the observed data between 20 and 60 keV) and inserting a single narrow Gaussian line at 67.87 keV at given flux level. The synthetic spectrum is then fit over the 20--75 keV bandpass to see if the Gaussian component is detected as described above.

We produce 1000 simulations for a each flux level using \texttt{fakeit} in XSPEC and determine how many of the simulations resulted in the detection of the line. We declare the flux level at which 50\% of the simulations produce detections to be the upper-limit. The upper limits vary spatially over the remnant (Figure \ref{fig:upperlimits_50}) due to the vignetting of the {\it NuSTAR} optics (i.e., the varying exposure as seen in Figure \ref{fig:TIboxes}, right panel) and the fact that the pointing strategy was optimized to cover the interior of the remnant with a uniform response.

\section{Discussion}
 \label{section:discussion}

\subsection{Understanding the relation between Ni/Fe and Ti}

Comparing the distributions of $^{44}$Ti and $^{56}$Ni is vital to understanding the physical conditions and kinematics of the innermost region of the supernova explosion. Most of the iron should result from the decay of $^{56}$Ni, so comparing the distributions of $^{44}$Ti and iron yields information on the initial distributions of nickel-rich ejecta and titanium-rich ejecta from the supernova explosion.

$^{44}$Ti is produced in a variety of nuclear processes, though the dominant process depends on the nature of the explosion (thermonuclear vs. core-collapse) and the structure of the star.

For thermonuclear supernovae, much of the $^{44}$Ti is formed in the burning of the He-shell or the C/O core.  If the density and temperature are sufficiently low (temperatures $<2-3 \times 10^9 K$ and densities $< 10^6 {\rm \,g\,\,cm^{-3}}$), the material does not burn all the way to $^{56}$Ni. Instead, the explosion produces $^{40}$Ca, $^{44}$Ti, and $^{48}$Cr \citep{Holcomb:2013et}.  In these conditions, $^{44}$Ti can be produced in regions where very little or no $^{56}$Ni is synthesized.

In contrast, for core-collapse supernovae like Cas A, the dominant $^{44}$Ti production occurs when the shock passes through the innermost silicon layer. The densities and temperatures are typically higher than those found in the $^{44}$Ti sites for thermonuclear supernovae with peak temperatures $>4 \times 10^9 K$ and densities $> 10^6 {\rm \,g\,\, cm^{-3}}$ \citep{Magkotsios:2010kr}.

The burning of silicon proceeds through photodisintegration when the rearrangement of the nuclei produces clusters of nuclei. This drives the material into nuclear statistical equilibrium, when the composition is determined by a balance between forward and reverse nuclear reaction rates. Depending upon the exact peak temperatures (and densities at these peak temperatures), \cite{Magkotsios:2010kr} identified 6 regions where different processes and reactions (e.g. different quasi-equilibrium clusters and
different nuclear statistical equilibrium freeze-out conditions) dictate the final nucleosynthetic yields. These
higher-density/hotter-temperature conditions mean that, in nearly all cases for core-collapse supernovae, at least some $^{56}$Ni is produced when $^{44}$Ti is produced. There are, however, scenarios in which the $^{56}$Ni / $^{44}$Ti ratio can fall to $\sim$100 and others where it is very high ($>10^{8}$).  Measuring this ratio (or the Fe / Ti ratio as a proxy) and its spatial variations can provide detailed clues into the nature of the explosion.

\subsubsection{Comparison of $^{44}$Ti and $^{56}$Ni ejecta in 3D}

\begin{figure}
\begin{center}
\includegraphics[width=1.0\columnwidth]{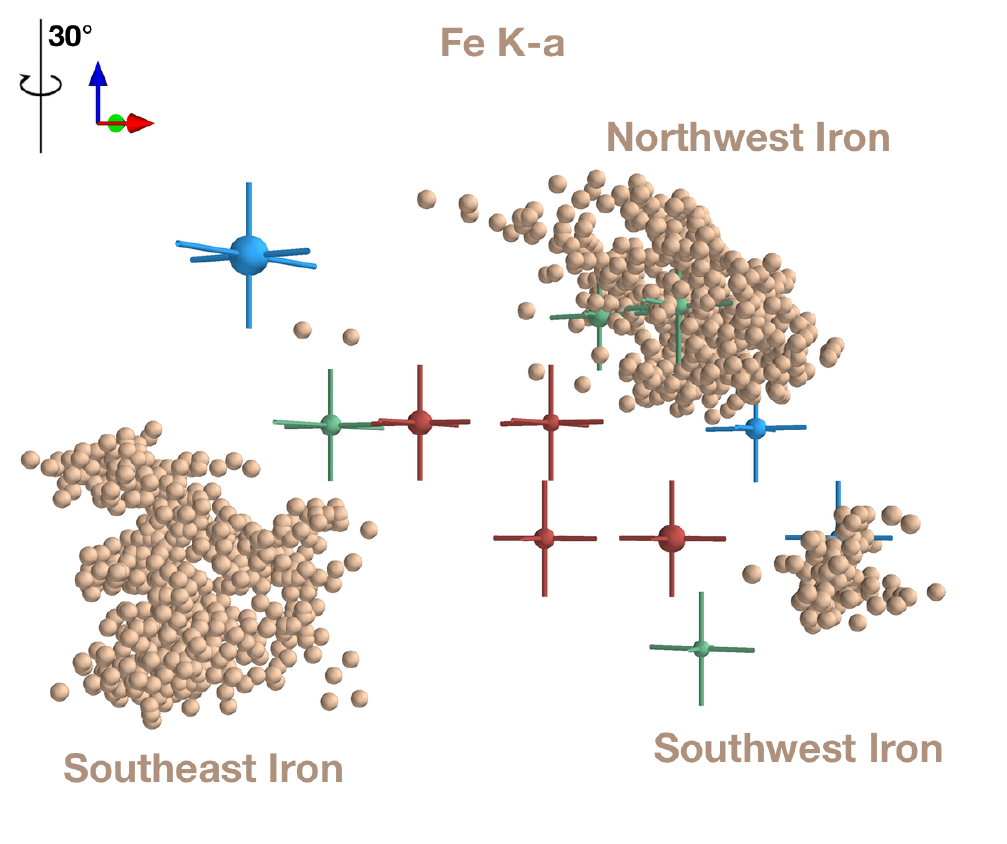}
\caption{\label{fig:ti_vs_fe}
A 3D comparison of the $^{44}$Ti ejecta observed by \textit{NuSTAR} and the iron emission observed by \textit{Chandra}. The $^{44}$Ti data points (here shown with 1-$\sigma$ error bars on all three dimensions) have the same color coding as in Figure \ref{fig:vectors_3d_grid} and are shown against the Fe-K data (grey-orange spheres) from {\em Chandra} \citep{DeLaney:2010fy}. The scene has been rotated by 30-degrees counter-clockwise around the blue (North) axis. The green and blue data points (indicating that the $^{44}$Ti ejecta is near or beyond the reverse shock) toward the Northwest and Southwest iron regions are consistent with the iron and titanium being co-located. Meanwhile, the blue data point to the top left (region 20b, projected towards the observer along the line of sight) is exterior to the reverse shock but lacks any iron counterpart. The red data points are interior to the reverse shock and may represent regions where there is diffuse, cold iron that has not yet been re-energized by the reverse shock and may be visible in the infrared. See the online journal for an animated version of this figure showing the full rotation about the center of the remnant.}
\end{center}
\end{figure}

Figure \ref{fig:ti_vs_fe} shows the comparison of the \textit{NuSTAR} $^{44}$Ti data and the Fe-K emission observed by \textit{Chandra}. The latter data are taken from \cite{DeLaney:2010fy} and as the iron is only visible in the X-rays after it has encountered the reverse shock we adopt the value derived by those authors to convert the line-of-sight velocity to distance of 0.032$''$ per km sec$^{-1}$ appropriate for the decelerated, reverse-shocked material.

In nearly all cases where we see $^{44}$Ti ejecta at or beyond the reverse shock (the green and blue data points in Figure \ref{fig:ti_vs_fe}) we also see emission from shocked iron (e.g., the regions labeled Northwest and Southwest Iron). The one exception to this is the blueshifted region 20b, which does have any obvious analog in the 3D distribution of shocked iron.

However, as we noted above, the line associated with region 20b is Doppler broadened beyond the nominal energy resolution of the detectors. This implies that the PSF of \textit{NuSTAR} is blending together several knots (or a shell) of  $^{44}$Ti ejecta rather than a resolving a single knot. In this case our conversion from the Doppler velocity to a 3D position may not be correct.

There is also some evidence for a trace amount of iron in the northern shell that is stationary with respect to the center of expansion of the remnant or slightly blueshifted toward the observer. It may be the case that there is a tenuous amount of iron that would be associated with region 20b but is too faint to observed when seen in projection against the (brighter) redshifted iron (i.e., the region labeled Northwest Iron in Figure \ref{fig:ti_vs_fe}). \\ 

% In general, it is not possible to produce $^{44}$Ti without also producing $^{56}$Ni, so it seems likely that the lack of iron emission associated with region 20b is an observational artifact.

%
%A subset of the Fe-K emitting regions described above are in the form of knots of ``pure" iron in Cas A that are characterized by their lack of associated silicon emission \citep{Hwang:2003dn,Hwang:2012gi}. The lack of observed emission from lighter elements suggests that these regions are associated with $\alpha$-rich freezeout during the explosive nucleosynthesis \citep{Hwang:2012gi} and these regions should, therefore, also contain $^{44}$Ti ejecta. However, as with the Fe-K emission, we find regions of pure iron emission with and without associated $^{44}$Ti ejecta (Figure \ref{fig:ti_vs_pure}). \\

\begin{figure}
\begin{center}
\includegraphics[width=1.0\columnwidth]{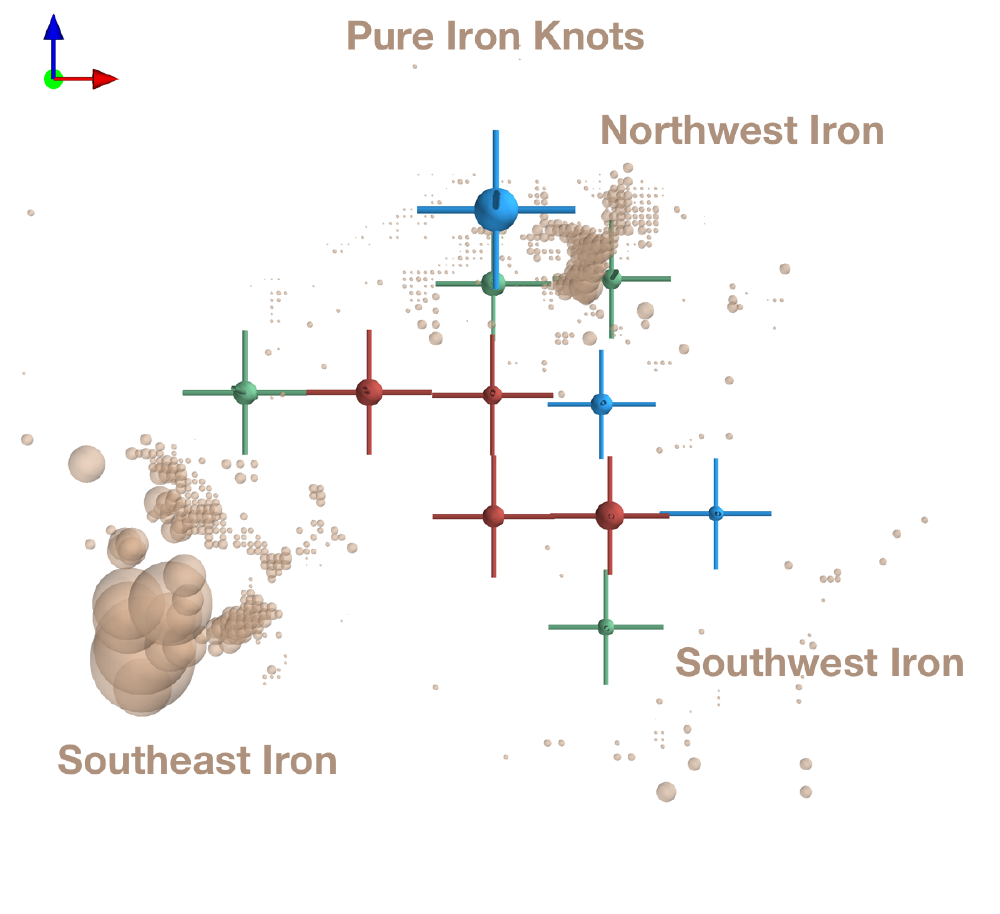}
\caption{\label{fig:ti_vs_pure}
A comparison of the 2D ``pure" iron map from \citet{Hwang:2012gi} and the $^{44}$Ti ejecta. The data and error bars are the same as shown as vectors in the middle panel of Figure \ref{fig:vectors_3d_grid} and are shown shown face-on as seen by the observer. The grey-orange spheres are the ``pure" iron ejecta and are scaled by the ejecta mass for each data point. These pure iron regions we expect to be associated with $\alpha$-rich freezeout and therefore also associated with the $^{44}$Ti ejecta. As for the total Fe K emission shown in Figure \ref{fig:ti_vs_fe}, there is pure iron in regions both with and without $^{44}$Ti ejecta. The Northwest, Southwest, and Southeast regions of iron are labeled for clarity. There is little, if any, pure ejecta in the Southwest where we otherwise find a good correlation between Fe-K emission and the $^{44}$Ti ejecta.}
\end{center}
\end{figure}

We remind the reader that we need to be careful when we interpret the observed distribution of iron. First, we only observe the iron that has been shock-heated (i.e., it has passed through the reverse shock), so X-ray measurements provide a partial observation of the iron produced in the supernova. Second, estimates of the iron mass based on X-ray measurements depend upon the excitation states of the iron and so will be affected by deviations from coronal equilibrium. Since the observed X-ray flux is proportional to the product of the iron mass and electron density, highly clumped material can also produce higher X-ray flux for the same iron mass than  for smoothly distributed material. Finally, if iron is present in the star and the circumstellar medium, the ejecta will contain ``swept-up" iron that cannot be distinguished from the iron synthesized in the explosion.

We can avoid the ambiguities in the ``swept-up" iron vs iron synthesized in the explosion if we only consider iron that was producing in the explosion. A subset of the \textit{Chandra} Fe-K emitting knots contain ``pure" iron; that is, the knots are characterized by a lack of associated silicon emission \citep{Hwang:2003dn,Hwang:2012gi}. The lack of observed emission from lighter elements suggests that these regions are associated with $\alpha$-rich freezeout during the explosive nucleosynthesis \citep{Hwang:2012gi} rather than incomplete silicon burning. We can thus attempt to quantify the relative production rates of $^{44}$Ti and iron by comparing the distribution of $^{44}$Ti observed by \textit{NuSTAR} with the distribution of pure iron observed by \textit{Chandra} (Figure \ref{fig:ti_vs_pure}).

\subsubsection{Beyond the reverse shock}
For the ejecta beyond the reverse shock, there is clearly some variation that produces a large yield of $^{44}$Ti in the Northwest and Southwest while suppressing the $^{44}$Ti in the Southeast. This is clear when comparing the $^{44}$Ti distribution to both the Fe-K 3D distribution (Figure \ref{fig:ti_vs_fe}) as well as to the pure iron distribution (Figure \ref{fig:ti_vs_pure}).

%The presence of the pure iron implies that the innermost ejecta were at such a high temperature and low density that incomplete silicon burning did not occur. By comparing the relative abundances of $^{44}$Ti and iron we can then infer the changes in the pressure and temperature conditions of the innermost ejecta during nucleosythensis. 

In the Northwest we integrate over the \textit{NuSTAR} regions 19, 20a, and 27 to recover a 68 keV line flux of 32.4 $\times 10^{-7}$ ph cm$^{-2}$ sec$^{-1}$, corresponding to an initial mass of $^{44}$Ti of 2.7 $\times 10^{-5}$ M$_{\odot}$. We similarly integrate over the regions in the \textit{Chandra} data that are at least 30$''$ North from the center of the remnant and to the West of the center of the remnant and find a pure iron mass of 0.014 M$_{\odot}$. This gives an Fe / Ti ratio of roughly 500.

We contrast this with the region of iron in the Southeast of the remnant. Integrating over all knots of pure iron ejecta from \textit{Chandra} that are South and East of the center of the remnant we find 0.018 M$_{\odot}$ of pure iron but no detectable $^{44}$Ti ejecta in the \textit{NuSTAR} data. For the regions that overlap with the southeast pure iron emission the upper limits on the 68 keV flux correspond to lower limits on the Fe / Ti ratio of $\sim$1000, or roughly twice the Fe / Ti ratio in the Northwest region. The total Fe mass in the southeast region far exceeds that of the pure iron ejecta and certainly some (or most) of this ejecta must have been synthesized in the explosion (i.e. via incomplete silicon burning, which would leave behind lighter elements to be observed) implying that the \textit{total} Fe / Ti ratio must be $\gg$1000 in this region.

\begin{figure}
\begin{center}
\includegraphics[width=1\columnwidth]{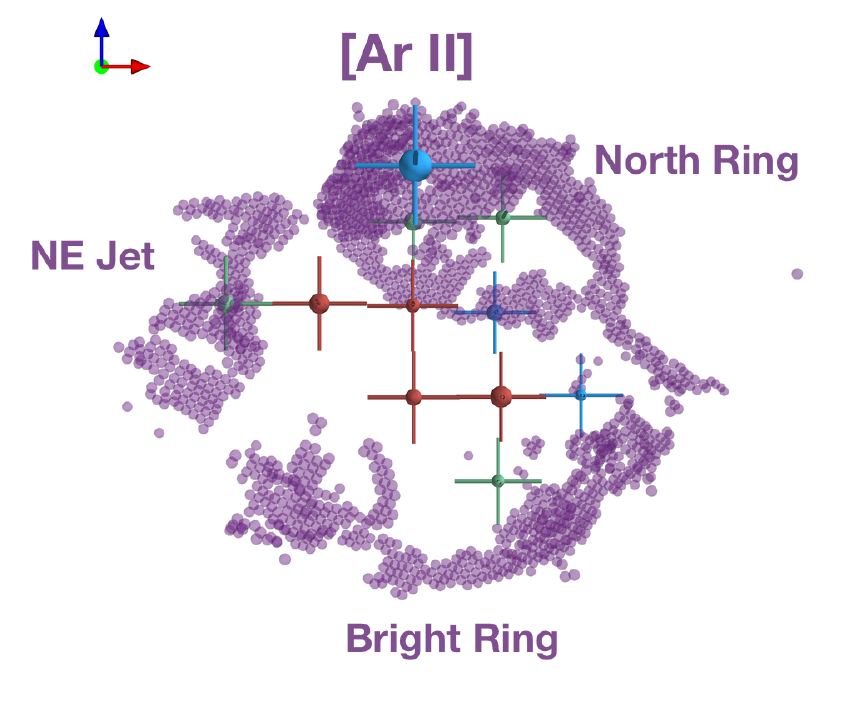}
\caption{\label{fig:errors_vs_ar}
The 3D distribution of the observed $^{44}$Ti ejecta compared with the [Ar II] emission observed by \textit{Spitzer} \citep{DeLaney:2010fy}. The $^{44}$Ti ejecta are shown with 1-$\sigma$ error bars and retain the color scheme from Figure \ref{fig:vectors_3d_grid}. The infrared data are shown in magenta along with labels to identify important features. The scene is displayed as seen by the observer, though an animation showing the full rotation of the remnant is available in the online journal. Upon rotation, the association of $^{44}$Ti ejecta with features in the [ArII] map (specifically in the ``Northern Ring" and the ring near the ``NE Jet") becomes more clear.}
\end{center}
\end{figure}

The difference in the yield of $^{44}$Ti may be a tracer of a change in the peak density of the innermost ejecta during nucleosynthesis. The yield of $^{56}$Ni (and therefore pure iron) is relatively insensitive to changes in density (and, indeed, we find a comparable mass of pure iron in the Northwest and Southeast), the $^{44}$Ti yield can depend sensitively on the density \citep[e.g.,][]{Magkotsios:2010kr,Magkotsios:2011cj}. The drop by roughly at least a factor two in the $^{44}$Ti yield between the Northwest and Southeast regions may be evidence for large-scale asymmetry in the peak density of the innermost ejecta in these directions.

\subsubsection{The unshocked interior}

There are knots of $^{44}$Ti emission interior to the Fe ejecta and the reverse shock (color-coded red). The combined flux from these regions of 4.5 $\times 10^{-6}$ ph~sec$^{-1}$~cm$^{-2}$, which corresponds to an initial mass of roughly 4$\times 10^{-5}$ M$_{\odot}$ (Figure \ref{fig:space_velocity_vs_flux}) or roughly a third of the total $^{44}$Ti mass in the remnant. This value is slightly underestimated because of the PSF-induced cross-talk between regions as discussed above, but is clear evidence for a significant fraction of $^{44}$Ti mass residing in the interior of the remnant.

\cite{Hwang:2012gi} predict 0.18-0.3 M$_{\odot}$ of unshocked ejecta in the interior of the remnant (roughly 10\% of the total ejecta mass), though it is not clear what fraction of this ejecta should be iron. If we make the assumption that the Fe / Ti ratio of $\sim$500 found in the Northwest then we estimate an unshocked mass of pure iron of $\sim$0.02 M$_{\odot}$ in the interior of the remnant. However, the uncertainties on this number are large; as we have seen in the exterior of the remnant there are large variations in the Fe / Ti ratio that are driven by changes in peak temperature and pressure of the innermost ejecta.

If $^{56}$Ni was produced in these regions then we might expect to observe iron emission in the infrared in the interior of the remnant. Such emission has not yet been observed \citep{Isensee:2010go, DeLaney:2014kj}, implying that either the interior iron ejecta are so diffuse that they cannot be detected, are in a higher ionization state due to photoionization from soft X-rays from the ejecta and thus cannot be observed by \textit{Spitzer}, or the ejecta are not present. Deeper observations to probe for a cold, diffuse source of iron are required to further constrain the Fe / Ti ratio in the interior of the remnant.

%
%Direct observational constraints of the Fe / Ti ratio in the interior of the remnant must await measurements of the unshocked iron mass from infrared observations searching for [Fe II], which has not yet been detected in the interior of the remnant in the \textit{Spitzer} IRS band, or in higher ionization states of iron with no strong IR lines.
%
% \\
%
%The un-shocked interior of the remnant contains over a third of the total $^{44}$Ti mass ($\sim$4 $\times 10^{-5}$ M$_{\odot}$). 
%

\begin{figure}[!hb]
\begin{center}
\includegraphics[width=1\columnwidth]{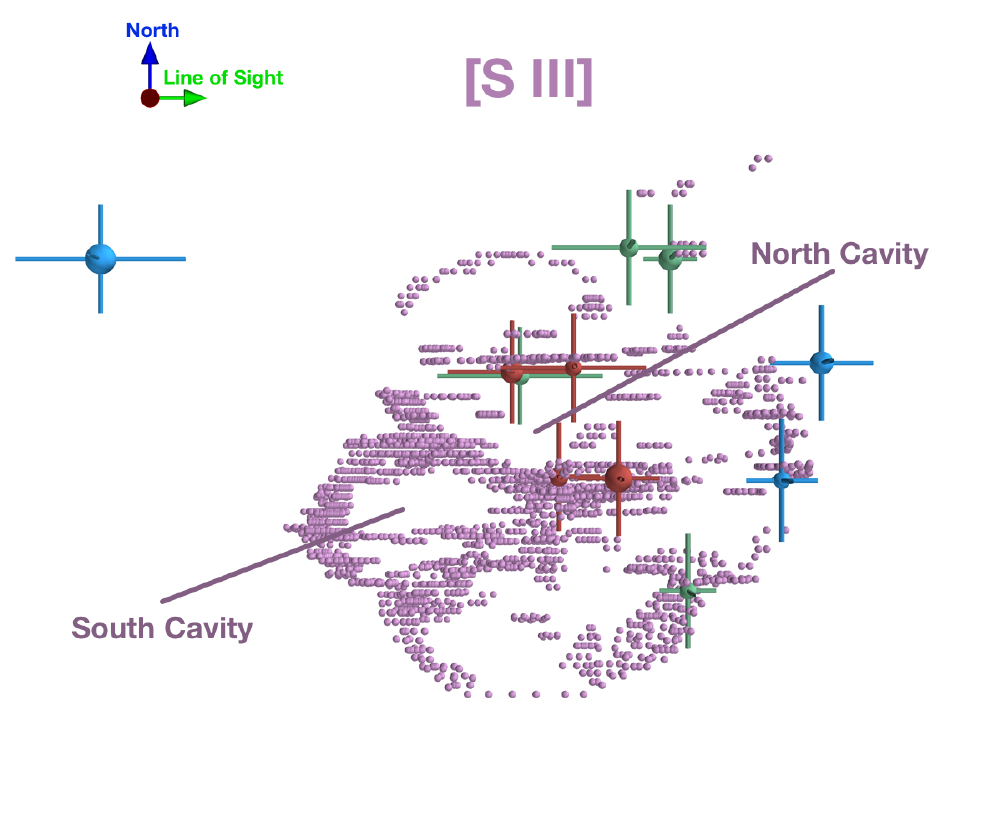}
\caption{\label{fig:ti_vs_s}
The 3D distribution of the observed $^{44}$Ti ejecta compared with the NIR [S III] emission \citep[data from][]{Milisavljevic:2015fw}. The $^{44}$Ti ejecta are shown with 1-$\sigma$ error bars and retain the color scheme from Figure \ref{fig:vectors_3d_grid}. The infrared data are shown in magenta along with lines to identify the north and south cavities. The scene is rotated by 90 degrees counterclockwise around the North axis and shows the view from the west (the same orientation as for Figure 2 from \citealt{Milisavljevic:2015fw}). Upon rotation, a rough association of the inner $^{44}$Ti ejecta with the north cavity  become more clear. We do not see any $^{44}$Ti ejecta associated with the blueshifted (i.e. the lower left side in the above rendering) South Cavity. An animated version of this figure showing the full rotation of the remnant is available in the online journal.}
\end{center}
\end{figure}

\subsection{$^{44}$Ti ejecta and infrared/optical features}

We can also compare the $^{44}$Ti ejecta with the emission seen at optical and infrared wavelengths. This again broadly falls into two categories: ejecta that have encountered the reverse shock and ejecta that are interior to the reverse shock.

For the shocked ejecta, we can use the [Ar II] 6.99 $\mu$m 3D maps from \textit{Spitzer} \citep{DeLaney:2010fy,Isensee:2010go}. Seen in the plane of the sky, the ejecta forms the feature known as the ``Bright Ring", while in 3D there are circular structures in the plane of the sky (labeled as the ``North Ring" and the ``NE Jet" structures, Figure \ref{fig:errors_vs_ar}).

We find that the $^{44}$Ti ejecta appear to correlate with both the North Ring and the NE jet (we recommend the movie available via the online journal for a more complete picture of these complex data). In the North Ring (where we also find Fe-K emission), this  could have been the result a bubble being blown in the material by the radioactive decay of clumps of $^{56}$Ni \citep[e.g.,][]{Li:1993eq,Blondin:2001if}. The North Ring happens to be coincident with the direction of one of the light echoes from Cas A, which showed that the photosphere of the supernova to the rear/northwest direction was moving faster than along the other lines of sight to the NE and SE \citep{Rest:2011fv}.

%We confirm our previous result that we do not see any additional emission along the jet/counter jet structure that extends in the northeast/southeast direction beyond the reverse shock. 

There is also emission from of unshocked ejecta seen in NIR [S III] line emission (906.9 and 953.1 nm) in the interior of the remnant \citep{Milisavljevic:2015fw}. These ejecta form bubble-like structures in the interior of the remnant (labeled the ``North" and ``South" cavities in Figure \ref{fig:ti_vs_s}). We find that the $^{44}$Ti ejecta may be associated with the northern cavity seen in the [S III] data, though we do not see any evidence for $^{44}$Ti associated with the southern, blueshifted cavity. We expect these bubbles to be associated with the decay of $^{56}$Ni and so we may be seeing variations in the resulting Fe / Ti ratio in the interior of the remnant similar to the variations that we observe beyond the reverse shock. 

\begin{figure}
\begin{center}
\includegraphics[width=1\columnwidth]{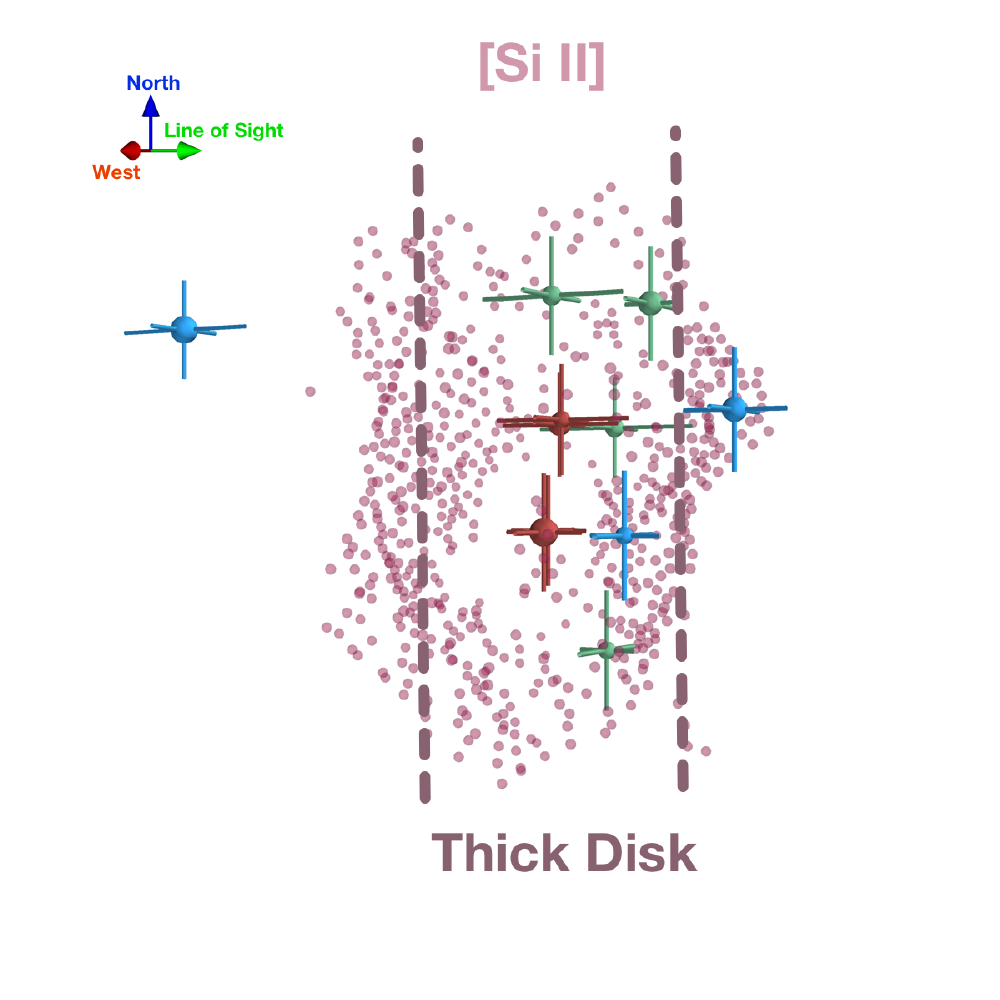}
\caption{\label{fig:ti_vs_si}
The 3D distribution of the observed $^{44}$Ti ejecta compared with the IR [Si II] emission observed by \textit{Spitzer} \citep{DeLaney:2010fy}. The $^{44}$Ti ejecta are shown with 1-$\sigma$ error bars and retain the color scheme from Figure \ref{fig:vectors_3d_grid}. The infrared data are shown in magenta along with lines to identify the extent of the ``thick disk" region. The scene is rotated by 120 degrees counterclockwise around the North axis. Upon rotation, the association of the inner $^{44}$Ti ejecta with the [Si II] map (specifically near the center of the thick disk) become more clear. We do not see any $^{44}$Ti ejecta associated with the blueshifted (i.e. the left side in the above rendering) face of the thick disk. An animated version of this figure showing the full rotation of the remnant is available in the online journal.}
\end{center}
\end{figure}

The interior unshocked ejecta is also seen in the infrared via [Si II] (34.8 $\mu$m) emission \citep[][]{DeLaney:2010fy}. These ejecta are apparently arranged into a ``tilted thick disk" (identified in Figure \ref{fig:ti_vs_si}) with a significant gap between the redshifted and blueshifted faces. We do not see any evidence for $^{44}$Ti ejecta associated with the blueshifted half of the thick disk, though we do see that the redshifted $^{44}$Ti ejecta are reasonably consistent with the red-shifted half of the thick disk. 

A future instrument with spatial resolution comparable to what is achieved by \textit{Chandra} or \textit{Spitzer} and spectral resolution better than \textit{NuSTAR} will likely be required to study the relative associations of the optical/infrared and $^{44}$Ti in the interior of the remnant in further detail.

\begin{figure*}
\begin{center}
\includegraphics[width=2\columnwidth]{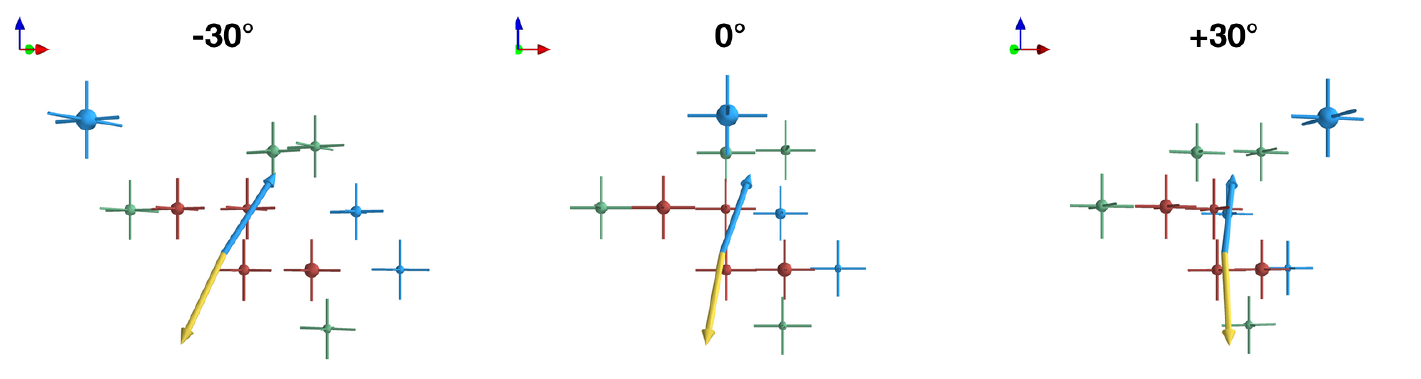}
\caption{\label{fig:vectors_3d_grid_with_cco}
The 3D distribution of the observed $^{44}$Ti ejecta compared with the CCO motion. The unit vectors are North (blue), West (red) and along the observer's line-of-sight (green). The data points show the $^{44}$Ti data along with 1-sigma errors and have the same color scheme as in Figure \ref{fig:vectors_3d_grid}. The blue vector shows the flux-weighted mean direction of motion of the $^{44}$Ti, while the gold vector shows the direction of motion of the CCO where the CCO is assumed to be moving opposite to the direction of the $^{44}$Ti along the line-of-sight and at a position angle of 169$^{\circ}$ degrees clockwise from North (see text). The center frame shows the remnant as seen by the observer, while the right/left frames have been rotated +/- 30$^{\circ}$ clockwise around North (blue) axis. An animation showing the full rotation of the remnant is available in the online journal. }
\vspace{12.8pt}
\end{center}
\end{figure*}

\subsection{$^{44}$Ti Ejecta and the CCO Kick}

The discovery of a point-like X-ray source (hereafter the central compact object, or CCO) in Cas A was one of the major results of the first high spatial resolution images of Cas A taken with \textit{Chandra} \citep{Tananbaum:1999ue}. The CCO is now thought to be consistent with a slowly rotating neutron star \citep[e.g.,][]{Chakrabarty_2001,Pavlov:2000jg,Mereghetti_2002} or, perhaps, a neutron star with a low surface magnetic field \citep[e.g.,][]{Pavlov:2009ida}.

If we accept that the CCO is a neutron star, then we can use it to study the natal kick of the neutron star \citep[e.g.][]{Burrows:1996eu,Burrows:2004tz,Wongwathanarat:2013fx}.  The CCO is offset from the center of expansion of the remnant by 7$''$.0 $\pm$0$''$.8 with a position angle of 169$^{\circ}$$\pm$8.4$^{\circ}$ in a 2004 epoch observation \citep{Fesen:2006caa}. For an explosion date of $\sim$1671 and a distance of 3.4 kpc, this corresponds to a plane-of-the-sky velocity of $\sim$330 km sec$^{-1}$. None of the major features in the remnant observed in the optical, infrared, or radio appear to match the CCO motion, though the bulk motion of the X-ray emitting ejecta in the remnant (210 km sec$^{-1}$ East and 680 km sec$^{-1}$ North, \cite{Hwang:2012gi}) is $\sim$150$^{\circ}$ from the direction of motion of the CCO.

The bulk of the $^{44}$Ti, however, is traveling in the west/northwest direction away from the observer. Using the flux-weighted mean of the 3D velocities given in Table \ref{tab:converted}, we find that the average direction of motion of the $^{44}$Ti ejecta has a position angle of $\sim$340$^{\circ}$ $\pm$15$^{\circ}$ (measured clockwise from Celestial North in the plane of the sky) and is tilted by 58$^{\circ}$ $\pm$ 20$^{\circ}$ into the plane of the sky away from the observer (Figure \ref{fig:vectors_3d_grid_with_cco}). The angle in the plane of the sky is almost exactly opposite to the inferred direction of the CCO motion.

If we assume that the 3D distribution of the $^{44}$Ti ejecta is a tracer of the 3D mass of the innermost ejecta at the time of the explosion and that the neutron star kick is related to the asymmetries in the ejecta mass distribution, then we can make a prediction about the velocity of the neutron star along the line of sight. We can then proceed assuming that there is a simple scaling relation between the total momentum of the $^{44}$Ti and the momentum of the neutron star:

\begin{equation}
C \times \vec{\rho}_{Ti} = \vec{\rho}_{NS}
\end{equation} 
where C is some proportionality constant that is roughly the ratio of the $^{44}$Ti ejecta mass to the total inner ejecta mass. Since we can directly measure the velocity components of both the $^{44}$Ti and the neutron star in the plane of the sky, we can fold the neutron star mass into the proportionality constant itself and make the comparison:
\begin{equation}
<\vec{v}_{Ti}>_{Sky} = C^{\prime} \times \vec{v}_{NS, Sky}
\end{equation} 
where $<\vec{v}_{Ti}>_{Sky}$ is the flux-weighted mean velocity of the $^{44}$Ti ejecta in the plane of the sky. The $^{44}$Ti has an observed plane of the sky velocity of 1450 $\pm$ 380 km sec$^{-1}$, so we cab compute the $C^{\prime}$ proportionality constant that will produce a plane of the sky velocity of the neutron star of 330 km sec$^{-1}$. Assuming the same proportionality constant applies to the line-of-sight direction we can convert the flux-weighted average of the $^{44}$Ti velocity along the line-of-sight (920 $\pm$ 510 km sec$^{-1}$) into an estimate for the line-of-sight velocity of the neutron star (205 $\pm$ 125 km sec$^{-1}$).

Unfortunately, testing this hypothesis is non-trivial, the uncertainties we quote here are large, and the physical scaling that we have performed here to convert between the $^{44}$Ti momentum and the neutron star kick is likely overly simplistic. However, the fact that the $^{44}$Ti ejecta ejecta are apparently moving in the direction opposite to that of the neutron star is highly suggestive that the two are related. When we also consider that the bulk ejecta is recoiling in roughly the same direction as the $^{44}$Ti and opposite to the direction of the neutron star \citep{Hwang:2012gi} and that the light echo in this region indicates that the exploding star was moving faster than along the other lines of sight to the NE and SE \citep{Rest:2011fv}, then we start to construct a more complete picture of the explosion. If the ejecta expands rapidly (perhaps as the result of a more energetic explosion), then the density can quickly drop into a region where $\alpha$-rich freezeout may occur, resulting in a high yield of $^{44}$Ti.

\subsection{Implications for Instabilities}

One of the most important unresolved issues currently facing the supernova simulation community is whether supernova explosions can be adequately modeled in 2D (i.e., the explosion can be described by axis-symmetric simulations) or whether they require 3D simulations to fully capture the relevant instabilities \citep[see e.g., recent reviews by][]{Fryer:2014dj,Janka:2016vc}. The favored interpretation for core-collapse supernova is that neutrino heating drives shock instabilities in the collapsing star \citep[e.g.,][]{Bethe:1985da}. These instabilities give rise to large spatial structures (i.e. those that can be described by ``low mode" spherical harmonics), or bubbles, in the ejecta that carry enough momentum to revive the stalled shock and explode the star. The {\em NuSTAR} 2D $^{44}$Ti map of Cas A strongly supports this low-mode convection model for the supernova engine \citep{Grefenstette:2014ds}.

We have also previously argued that even the 2D images of the $^{44}$Ti ejecta suggest that large scale structures dominate the ejecta distribution rather than small turbulent eddies (i.e., features that can be described by ``high-mode" spherical harmonics) or ``jet"-like features that can result from the collapse of a rapidly rotating massive star like those that are present in Type Ib/Ic supernovae and/or gamma-ray bursts \citep[e.g.,][]{Mosta:2015fb}. The generation of these large structures in 3D may be related to the Standing mode Accretion Shock Instability (SASI), which redistributes power to lower spherical harmonics in 3D simulations while turbulence will drive power to higher order modes \citep{Janka:2016vc}. This can also occur in Rayleigh-Taylor driven convection \citep[e.g.][]{Herant_1995}. 

The fact that we now see large, coherent structures in the 3D distribution of the ejecta is further evidence that the large spatial instabilities do not cascade down to small spatial scales in less than a dynamical timescale. This is especially true when we consider the spatial variations of the measured Fe / Ti abundance, which is nearly bipolar in structure and may be the best tracer for density asymmetries in the innermost ejecta during explosive nucleosynthesis. 

\section{Summary}

We have presented results from the 2.4 Ms \textit{NuSTAR} campaign designed to study the $^{44}$Ti ejecta in Cas A. These data provide the first opportunity to study the 3D distribution of $^{44}$Ti in Cas A. The ability to spatially resolve the emission from the $^{44}$Ti ejecta provides us with a new probe for studying nucleosynthesis in the supernova explosion by studying the relative spatial distributions of the $^{44}$Ti-rich ejecta and the $^{56}$Ni-rich ejecta.

The average momentum (i.e., the flux-weighted average of the $^{44}$Ti ejecta velocities) gives a resulting vector rotated in the plane of the sky by $\sim$340$^{\circ}$ $\pm$ 15$^{\circ}$ (measured clockwise from Celestial North) and tilted by 58$^{\circ}$ $\pm$ 20$^{\circ}$ into the plane of the sky away from the observer. The plane-of-the-sky velocity is almost precisely opposite the direction of the Cas A CCO. This is highly suggestive that the $^{44}$Ti ejecta is tracing out the instabilities that led to the neutron star kick in Cas A. We therefore expect that the neutron star should have a significant transverse (line-of-sight) velocity towards the observer, though we have no observational means of testing this hypothesis.

We find $^{44}$Ti ejecta interior to the reverse shock, though these ejecta cannot be definitively associated with known features observed in the optical or the infrared. The present-day flux from this ejecta implies that there is an initial mass of $\sim$4 $\times 10^{-5}$M$_{\odot}$ of $^{44}$Ti interior to the reverse shock. If we assume this interior ejecta has a comparable Ni / Ti ratio to the regions exterior to the reverse shock (implying an Fe / Ti ratio of $\sim$500) then we estimate that there is 0.02 M$_{\odot}$ of ``hidden"  iron in the interior of Cas A, though we caution that this value is highly model dependent. 

Where we see $^{44}$Ti ejecta near or exterior to the reverse shock in 3D we generally see emission from shock-heated iron, which should mostly be descended from $^{56}$Ni that is synthesized along with the $^{44}$Ti in the explosion. This is true both of iron that is associated with lighter elements which may have the result of incomplete silicon burning as well as regions of ``pure" iron that we think result from $\alpha$-rich freeze-out. While there is some evidence for $^{44}$Ti ejecta exterior to the reverse shock where we do not observe any associated iron we are not convinced that either the interpretation of the 3D location of Doppler-broadened region of $^{44}$Ti ejecta is correct or that the lack of observed iron implies that the iron is not present.

Conversely, we do find regions of iron emission exterior to the reverse shock where we do not see associated $^{44}$Ti emission. This is true both for the iron we think is associated with incomplete silicon burning and the iron we think is associated with $\alpha$-rich freezeout. The upper limits on the presence of $^{44}$Ti in these exterior regions suggests that the $^{44}$Ti yield must be suppressed by at least a factor of two relative to the yield of $^{56}$Ni in these regions to explain the lack of observed $^{44}$Ti ejecta.

  \section*{Acknowledgments}

We would like thank Dan Milisavljevic for providing the [S III] data files, as well as Thomas Janka, Raph Hix, and Adam Burrows for their helpful comments. This work was supported under NASA contract NNG08FD60C and made use of data from the {\it NuSTAR} mission, a project led by the California Institute of Technology, managed by the Jet Propulsion Laboratory, and funded by NASA. JML was supported by the NASA ADAP grant NNH16AC24I.

We thank the {\it NuSTAR} Operations, Software and Calibration teams for support with the execution and analysis of these observations. This research made use of the {\it NuSTAR} Data Analysis Software (NuSTARDAS), jointly developed by the ASI Science Data Center (ASDC, Italy) and the California Institute of Technology (USA). This research also made extensive use of the IDL Astronomy Library (http://idlastro.gsfc.nasa.gov/). Additional figures were produced using the Veusz plotting package (\copyright~2003-2016 Jeremy Sanders). 3D figures and movies were produced via the Anaconda Software Distribution (https://www.continuum.io) of python and mayavi2 \citep{ramachandran2011mayavi}. \\

{\em Facilities}: {\em NuSTAR}, {\em Chandra}, \textit{Spitzer}

\bibliography{casA_paper2_bibliography}

\end{document}